\documentclass[journal]{IEEEtran}
\usepackage{bm}
\usepackage{cite}
\usepackage[pdftex]{graphicx}
\usepackage{xcolor}

\newcommand{\sm}[1]{\textcolor{black}{#1}}

\usepackage{amsmath}
\usepackage{amssymb}
\usepackage{blkarray}

\usepackage{subcaption}

\begin{document}
%

\title{A Graph-based Molecular Communications Model Analysis of the Human Gut Bacteriome} 

\author{Samitha~Somathilaka,
        Daniel~P.~Martins,
        Wiley Barton, Orla O'Sullivan, Paul~D.~Cotter,~Sasitharan~Balasubramaniam,
\thanks{S. Somathilaka, Daniel P. Martins are with VistaMilk Research Centre and the Walton Institute for Information and Communication Systems Science, Waterford Institute of Technology, Waterford, X91 P20H, Ireland. E-mail: {samitha.somathilaka,daniel.martins}@waltoninstitute.ie.}
\thanks{S. Balasubramaniam is with the Department of Computer Science and Engineering, University of Nebraska-Lincoln, 104 Schorr Center, 1100 T Street, Lincoln, NE, 68588-0150, USA. E-mail:sasitharanb@gmail.com}
\thanks{Wiley Barton, Orla O'Sullivan and Paul D. Cotter are with VistaMilk Research Centre and the Teagasc, Food Research Centre, Moorepark, Ireland, P61 C996. E-mail: paul.cotter@teagasc.ie}}

\maketitle

\begin{abstract}
Alterations in the human gut bacteriome can be associated with human health issues, such as type-2 diabetes and cardiovascular disease. Both external and internal factors can drive changes in the composition and in the interactions of the human gut bacteriome, impacting negatively on the host cells. 
In this paper, we focus on the human gut bacteriome metabolism and we propose a two-layer network system to investigate its dynamics. Furthermore, we develop an \textit{in-silico} simulation model (virtual GB), allowing us to study the impact of the metabolite exchange through molecular communications in the human gut bacteriome network system. Our results show that the regulation of molecular inputs can strongly affect bacterial population growth and create an unbalanced network, as shown by the shift in the node weights based on the molecular signals that are produced. Additionally, we show that the metabolite molecular communication production is greatly affected when directly manipulating the composition of the human gut bacteriome network in the virtual GB. These results indicate that our human GB interaction model can help to identify hidden behaviors of the human gut bacteriome depending on the molecular signal interactions. Moreover, the virtual GB can support the research and development of novel medical treatments based on the accurate control of bacterial growth and exchange of metabolites.  
\end{abstract}


%
\IEEEpeerreviewmaketitle

\section{Introduction}


\IEEEPARstart{T}{he} Gut Bacteriome (GB) is an ecosystem of a massive number of bacterial cells which play a vital role in maintaining the stability of the host's metabolism\cite{ursell2012defining}.
The bacterial populations of the GB build complex interaction networks by exchanging metabolites with the host and/or other bacterial populations, resulting in the production of new metabolites or other molecules, such as Short Chain Fatty Acids  (SCFA) (essential for the host's health) and proteins \cite{sanna2019causal}. This process occurs inside of each bacterial cell and is supported by specific complex gene regulatory networks, which are modulated depending on the established interaction networks in the human GB. 

External factors such as the availability of nutrients, antibiotics and pathogens can affect this interaction network resulting in disruptions  to the overall composition and, in turn, behavior of the human GB \cite{wen2017factors}. These factors mainly alter the compositional balance of the human GB, i.e. dysbiosis, disrupting the metabolite production  \cite{iljazovic2020perturbation}. In humans, these GB changes have a significant impact on the host’s health and may lead to many diseases including \sm{inflammatory} bowel disease, type-2 diabetes, and obesity. Therefore, several studies have been undertaken to precisely identify  the causes for microbial behavioral alterations and their consequent health effects in humans and animals \cite{gupta2020predictive,lynch2019translating,He2018}. For example, Yang et al. \cite{yang2020landscapes} performed a cross-sectional whole-genome shotgun metagenomics analysis of the microbiome and proposed a combinatorial marker panel to demarcate microbiome related major depressive disorders from a healthy microbiome. From a different perspective, Kim et al. introduced a split graph model to denote the composition and interactions of a given human gut microbiome \cite{kim2019novel}. They studied three different sample types (classified as healthy or Crohn's disease microbiome) to analyse the influence of microbial compositions on different behaviors of the host's cells. Inspired by these works, we propose a novel tool to further characterize the interactions among the bacterial populations often found in the human GB.

In this paper we propose a two-layer interaction model supported by the exchange of molecular signals, i.e. metabolites, to model the human GB. 
Here, we identify the interactions between bacterial cells as \textbf{Molecular Communications} (MC) systems and their collective behavior as a MC network. MC aims to model the communication between biological components using molecules as information \cite{akyildiz2019moving} and it is fundamental to characterize the exchange of metabolites in our two-layer interaction model.

In the graph network, bacterial populations act as nodes while the edges represent the interactions between them. This interpretation allows to quantify the behaviors of the human GB using graph theoretical incorporating MC analysis to understand impacts from distances between different graph states and variations of node/edge weights. The theoretical quantification of the node behaviors and edge behaviors that will be discussed in subsequent sections explain the quantification of the graph behaviors using Flux Balance Analysis (FBA).
\sm{Moreover, conducting \textit{in-vivo} or \textit{in-vitro} experiments on the human GB to extract data related to each interaction of the network often requires a significant amount of resources.}
Hence, we designed an agent-based simulator (henceforth  named  as  virtual  GB) to simulate the human GB which produces data that can be used to quantify the same set of measures by avoiding complex calculations. The main reason for the complexity of theoretical calculations using FBA is the number of parameters and their stochastic nature. The virtual GB performs the behaviors of the human GB considering natural characteristics. Hence, the generated data represents bacterial behaviors that are influenced by the aforementioned stochastic parameters.


\begin{figure*}

     \centering
     \begin{subfigure}[b]{0.55\textwidth}
         \centering
         
         \includegraphics[width=\textwidth]{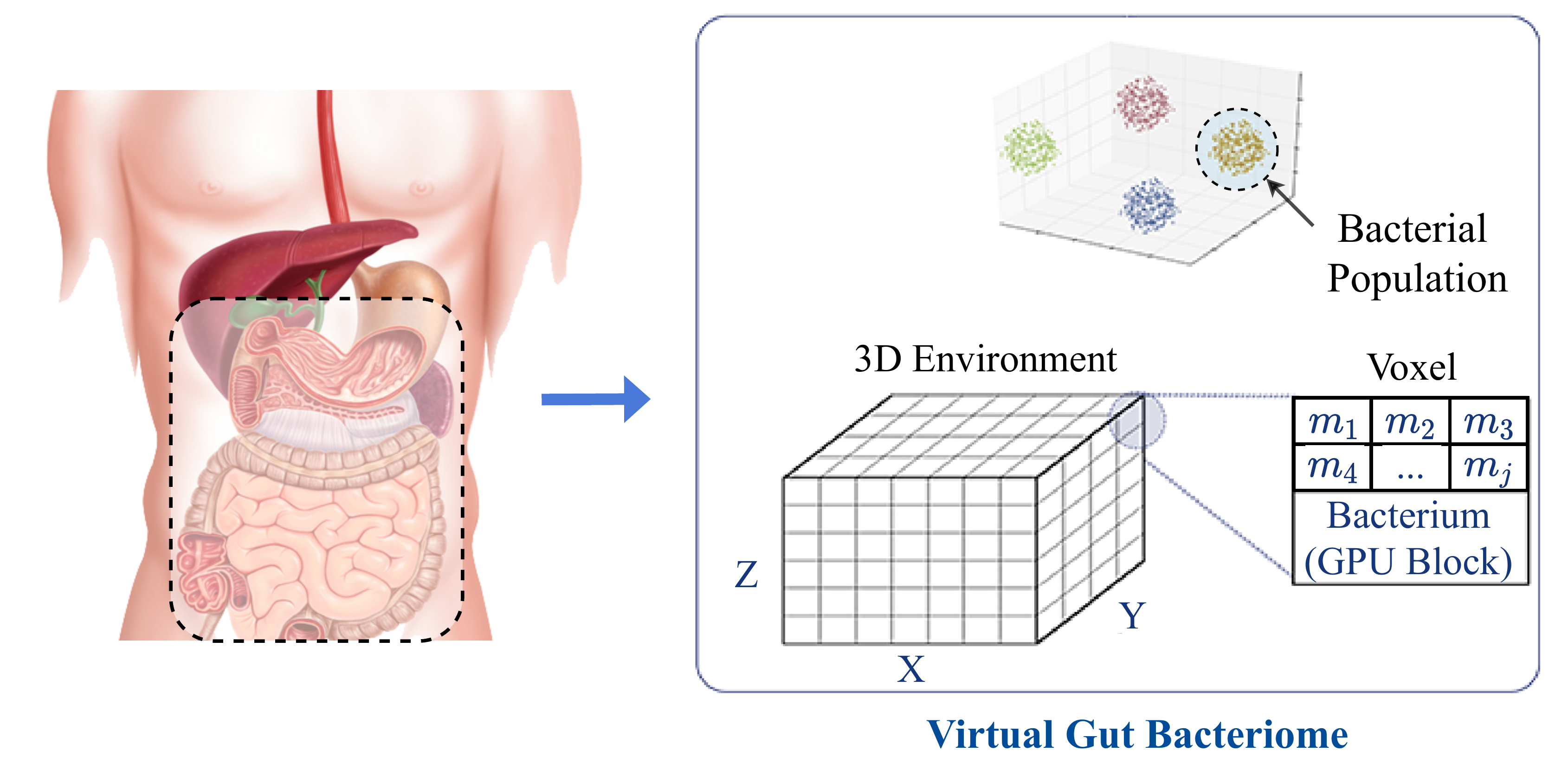}
         \caption{}
         \vspace{63pt}
         \label{fig:VGB}
     \end{subfigure}
     \begin{subfigure}[b]{0.43\textwidth}
         \centering
         \includegraphics[width=\textwidth]{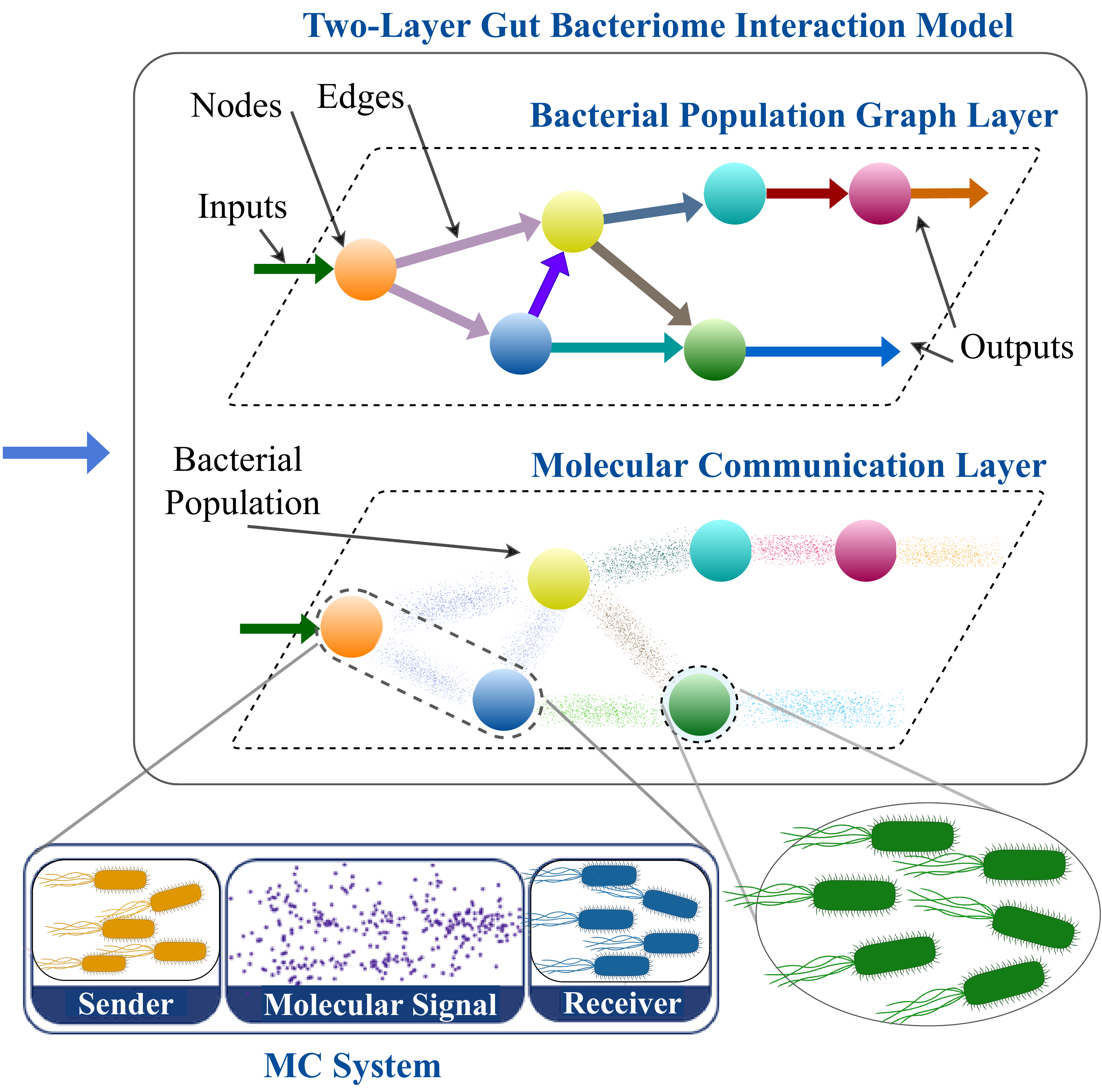}
         \caption{}
         \label{fig:TwoLayerModel}
     \end{subfigure}
        \caption{Illustration of the system model. (a) We recreated the human GB functionalities on virtual GB using voxel architecture and parallel processing dedicating one GPU block for each bacterial cell to produces quantitative data on MC layer, and (b) we propose a two-layer system model to investigate the molecular interactions simulated in the virtual GB.}
        \label{fig:SystemModel}
\end{figure*}

Our main contributions are as follows:
\begin{itemize}

 \item \textbf{Design of a two-layer interaction model of the human GB}: The gut bacteria
    consume, metabolise and secrete metabolites as single cells and use them to form a complex interaction network among the different bacterial populations in the human GB. Hence, in this study we design a layered interaction model to investigate the dynamics of the human GB based on the exchange of metabolites.
    \item \textbf {Analysing molecular communication impact on the human GB graph structure:}
    Deviations of bacterial populations' metabolism cause alterations in molecular interaction within the human GB, which  may impact the graph layer structure. We analyse this relationship between the MC measures and the graph structure of the human GB in terms of graph nodes and edges behaviours.
    \item \textbf{Development of a human GB simulator to perform \textit{in-silico} experiments:}
    We design and utilize an \textit{in-silico simulation model} of the human GB to investigate the direct and hidden interactions among bacterial populations based on the exchange of metabolites.
    
    \item \textbf {Explaining the dynamics of human GB using graph analyses:}
    The bacterial growth and changes in the human GB metabolism can be interpreted as a result of a series of cascading metabolic activities through bacterial interactions. 
    In this study, we explain the dynamics of bacterial growth and metabolism changes using graph analysis on the bacterial interactions.

\end{itemize}

In the next sections, we detail our approach to model the human GB and assess its network performance. Section \ref{sec:background}, we describe the basics of the human GB, and highlight the existent gaps that this research aims to address. Our proposed model is detailed in Section \ref{sec:two-layer}. Then, in Section \ref{sec:VirtualGB}, we introduce the simulation environment built to utilize metagenomics data and perform \textit{in-silico} experiments with the human GB. The metrics considered in this paper are introduced in Section \ref{sec:SystemDynamics}, and our analysis results are presented in Section \ref{sec:AnalyticalResults}. Finally, our conclusions are shown in Section \ref{sec:conclusions}. 






\section{Background on the Human GB Model}\label{sec:background}

The human GB is the bacterial ecosystem residing inside the human digestive system, comprising of approximately 1000 species interacting with each other, and carrying out crucial functions such as nutrient metabolism and immunomodulation of the host \cite{jandhyala2015role}. These bacteria do not manifest their cellular functions as individual entities but exhibit various social behaviors such as commensalism \cite{hooper2001commensal}, amensalism \cite{garcia2017microbial}, mutualism\cite{chassard2006h2}, parasitism and competition by interacting with other populations mainly using molecules (e.g., proteins, metabolites and \textit{quorum} sensing) \cite{hasan2015social}. Moreover, similar bidirectional exchanges of molecules occur between the GB bacterial populations and the human gut cells. These bacteria utilize products of the host metabolism activities or dietary components from the gastrointestinal tract to convert into various products essential for the host, through different metabolic pathways \cite{visconti2019interplay}.

This composition of human GB is a crucial driver for processing of metabolites (i.e., small molecules produced and used in metabolic reactions) in the lower intestine, which significantly impacts the health of the host \cite{singh2017influence}. Hence, any imbalance of the GB may result in negative impacts on the human health ranging from metabolic deficiencies to diseases such as type-2 diabetes, inflammatory bowel diseases and cancers \cite{henson2017microbiota,bjerrum2015metabonomics}. The composition of the human GB differs between individuals, and it depends on various factors including dietary patterns, gut diseases, exercise regimes, antibiotic usage, age, and genetic profiles \cite{rajoka2017interaction}.  
\begin{figure}[t!]
\centering
\includegraphics[width=\columnwidth]{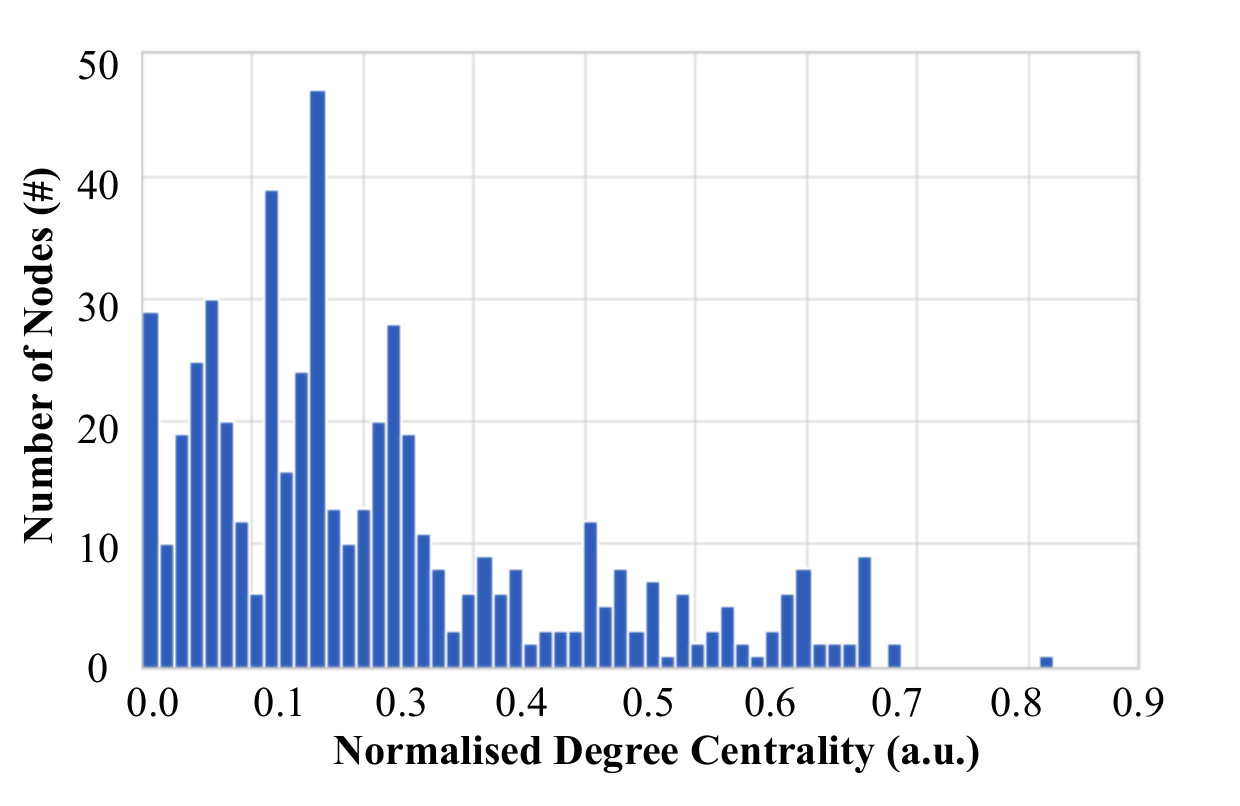}
\caption{Degree Centrality distribution of the full human GB MC network.}
\label{fig:FullGraph_DC}
\end{figure}

The relationship between the GB and the human gut not only benefits the host, but also the bacterial populations. They utilize the nutrient availability (i.e. the metabolic inputs to the GB) to modulate their growth and improving their survivability \cite{shanahan2017feeding}. Some bacteria can shift their metabolic pathways based on the extracellular signals such as the concentrations of nutrients and the growth condition \cite{shimizu2013metabolic}. For instance, \textit{Roseburia inulinivorans} switches the gene regulation switches between consumption of glucose and fucose in SCFA production according to the nutrient availability\cite{scott2006whole}. In addition to that, some of the bacterial populations will be highly benefited than others depending on the metabolite/nutrient that is being processed by the GB, which will require different signalling pathways, resulting in single or multiple bacterial interactions. For example \textit{Eubacterium rectale} consumes acetate as a growth substance and produces butyrate which is one of the most important SCFA \cite{riviere2016bifidobacteria}\cite{o2016bifidobacteria}. Similarly, the growths of \textit{Roseburia intestinalis} and \textit{Faecalibacterium prausnitzii} are also stimulated by acetate \cite{rowland2018gut}. Hence, it is evident that overall metabolic functionality relies on the composition of the GB and the metabolic inputs have a significant impact on GB composition.

The bacterial population signalling process in the GB is quite similar to routing and relaying information in a conventional network system, and has inspired different network models (including ours) of the human GB interactions. For example \cite{naqvi2010network} used a network-based approach to characterize the human gut microbiome composition and analyzed the healthy vs diseased states using network statistics. Another study focuses on the use of Boolean dynamic models that combines genome-scale metabolic networks, to determine the metabolic deviations between community members, which was used to characterize their metabolic roles of interactions\cite{steinway2015inference}.

\section{Two-layer Human GB Interaction Model}\label{sec:two-layer}

In this paper, we represent the metabolic interactions between bacterial populations in the human GB as a two-layer interaction model.
Figure \ref{fig:SystemModel} explains how we designed the model. First, the compositional and behavioral data on the human GB is extracted from the databases and literature. Then, the extracted data is used in implementing the virtual GB (Figure \ref{fig:VGB}). Finally, the virtual GB simulates the human GB functionalities according to various experimental setups (later explained in section \ref{sec:AnalyticalResults}) and produces data on bacterial, molecular and gut environmental behaviors. Then the produced data is analysed according to the introduced Two-layer interaction model as shown in Figure \ref{fig:TwoLayerModel}.

\begin{figure*}[t!]
\centering
\includegraphics[width=\textwidth]{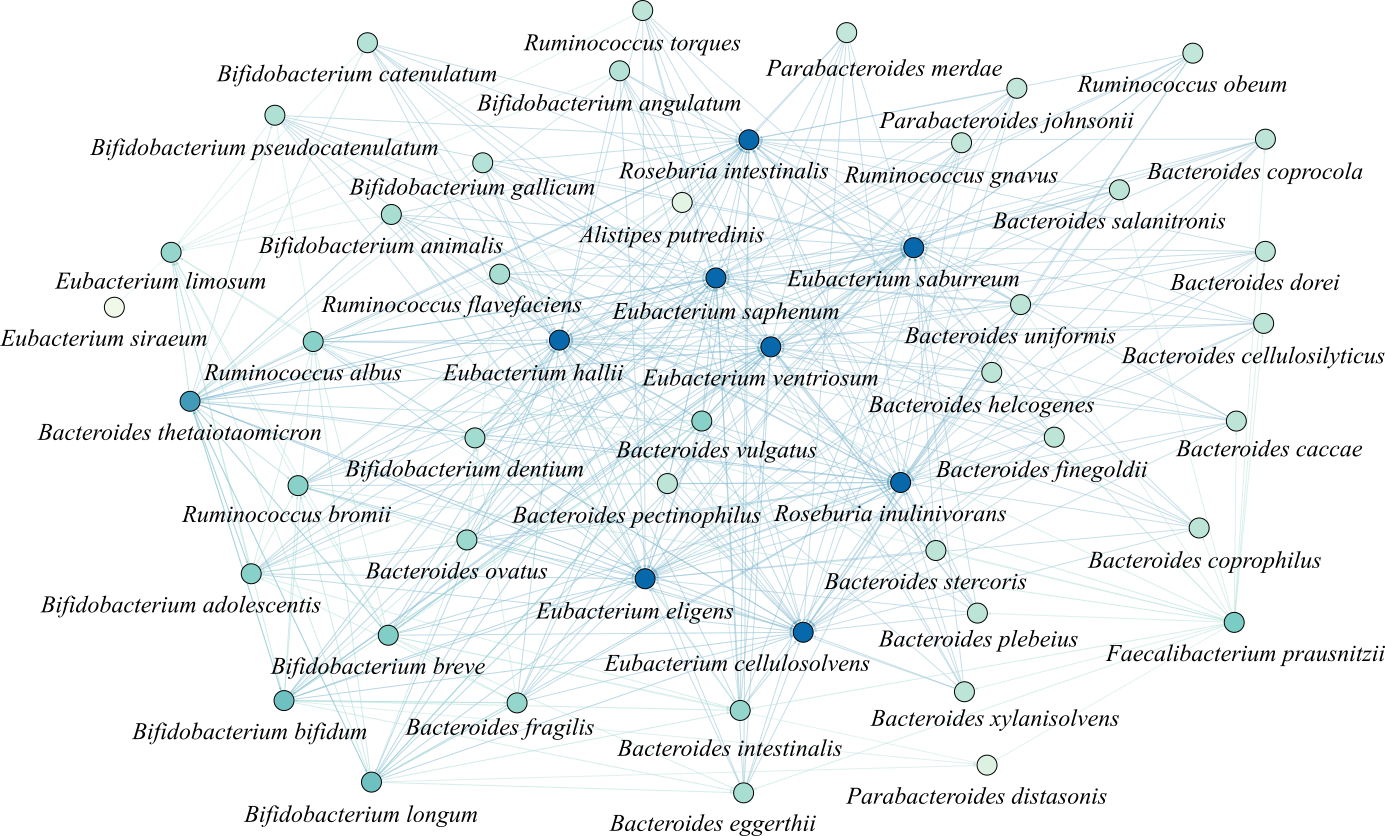}
\caption{Illustrates a subgraph of the human GB only considering species of nine genera related to SCFA. The nodes are color-coded according the degree ranking, where the darker color indicates higher number of inward and outward interactions while nodes with lesser number of interactions are with lighter color.}
\label{fig:SubGraph}
\end{figure*}

The upper layer of this model which is the Bacterial population graph layer defines the interconnections and overall structure of the human GB where we model the bacterial populations and host as nodes, and interactions between them as edges. To minimize the complexity of the model, this graph layer considers genera as nodes as species in the same genus share a common ancestral origin and the data availability. Further, the edges of the network represent the direct connections between the nodes that produce a particular metabolite and the nodes that consume the corresponding metabolite (An example graph network of the human GB can be seen in Figure \ref{fig:SubGraph}). 
In this layer, we can observe which bacterial genus is predominant in the human GB to measure the compositional changes and its impacts on the host's health. In other words, this model allows us to investigate the network topology of the human GB.

The bottom layer consists of the cascaded molecular communications systems created by the bacterial populations to establish their exchange of metabolites and support their network structure. Here, each node is viewed as a molecular transceiver and the edges are the communications channels interconnecting the nodes. Furthermore, this model extends to the molecular signals that reach the human GB from the environment, as well as, the ones that are output from the human GB and return to the environment. The interactions represented in this layer are dynamic and will depend on several environmental conditions, such as media characteristics, bacterial population sizes, and human GB composition. Please note, this is the layer where we initially observe the impacts of any alterations on the human GB composition (we further model and analyse this effect in Section \ref{sec: InputVsStructure}). The upper layer and the bottom layer are further described in the following sections.

\subsection{Upper Layer - Bacterial Population Graph Layer}\label{sec:upper_layer}
Bacteria display a wide variety of social behaviors, and this can lead to processes such as metabolism of molecules or coordinated biofilm formation \cite{unluturk2016impact}. The bacteria's ability to consume and produce multiple metabolites results in dense  interaction patterns that can lead to challenges in the analysis. However, representing the interactions of human GB as a graph network can provide a new avenue towards analyzing the bacterial communication and the impact on the overall bacteriome stability. 

Our human GB interaction model aims to provide a better global view of the functionality of the human GB, leading to the understanding of the causes and effects of its imbalance to propose precise alterations to fix such issues. Therefore, we model the human GB as follows. We first consider that all bacterial cells $b^{B_{k}}$ of a bacterial population $B_{k}$ (where $k$ is the bacterial population identifier) perform the same series of metabolic functions to process the metabolites input in the human gut. Each node of the proposed MC network is a bacterial population and comprises the collective metabolic functions of all cells. \sm{Let $\Omega$ be the set of all agents we considered in our human GB interactions study, i.e. host cells and bacterial populations, $\Omega=\{host,B_k\}$. In this case, the molecular intake of population $B_{k}$ from $\Omega$, $C_{(\Omega,B_k)}$. is considered $C_{(\Omega, B_k)}\simeq\sum c_{(\Omega, b^{B_{k}})}$ where $C$ represents population interactions, $c$ represents the intercellular interactions and $c_{(\Omega, b^{B_{k}})}$ is the molecular reception of bacterial cell $b^{B_{k}}$ from a $B_{k}$ source. In the same way, molecular emission of the population is considered the combined molecular emission of all bacterial cells of the particular population, $C_{(B_k, \Omega)}\simeq\sum c_{(b^{B_{k}}, \Omega)}$, where  $C_{(B_k,\Omega)}$ is the molecular emission from population 
$B_k$ to any receiver (host or other bacterial population) and $c_{(b^{B_{k}}, \Omega)}$ is the molecular emission of single bacterial cell of the population $B_k$ to any receiver. Additionally, the metabolite consumed by the bacterial cell $b^{B_k}$,  $M_{Con}({b^{B_{k}}})$ is obtained as $M_{Con}(b^{B_{k}})=c_{(\Omega, b^{B_k})}-c_{(b^{B_k}, \Omega)}$. Hence the metabolite consumption of a bacterial population is defined as $M_{Con}(B_{k}) \simeq \sum M_{Con}(b^{B_{k}})$.}

\sm{Next, we map the interactions between bacterial populations to a directed multi-graph network $\Gamma = (B, C, B^{s}, B^{d}, M)$ where $B$ is the set of all bacterial populations, $C$ is the set of all interaction in the human GB, $B^{s}\in B$ is \sm{the bacterial population} interaction sources, $B^{d}\in B$ \sm{the bacterial population} interaction destinations and $M$ is the set of metabolites. Using this definition and the data from NJS16 database, we can create the full graph network of the human GB $\Gamma_{full}$ where $|B|=532$, $|C|=30,085$ \cite{sung2017global}. Figure \ref{fig:FullGraph_DC} shows the normalized degree distribution of the full network $\Gamma_{full}$ which aligns with a power-law $P_{\text{deg}}(\kappa) \propto \kappa^{-\gamma}$ (where $\kappa$ is the degree and $\gamma$ is the decaying factor) distribution. Hence, we categorize the human GB as a scale-free network. Additionally, we emphasize a dominating quality of scale-free network which is the relative nodes that are common with a high number of interactions. Nodes with a significant number of edges are considered as hubs that perform specific functions. In this work we reduce such complex MC network of the human GB by focusing our model on the bacterial population interactions related to one of the main metabolic interactions, i.e. production of SCFAs. For example, Figure \ref{fig:SubGraph} represents a bacterial population subgraph $\Gamma_{sub}$ of the human GB, where the number of bacterial populations $|B|=50$, number of interactions $|C|=393$ and number of molecules used in interactions $|M|=230$, which will be investigated in the further sections of this work. In this graph, some nodes are involved in a significantly large number of interactions, compared to others. The color scheme used for the nodes represents the degree ranking, where darker colors indicates the higher degree ranking while lighter colors indicates the opposite.}



\begin{figure}[t!]
    \centering
    \includegraphics[width=\columnwidth]{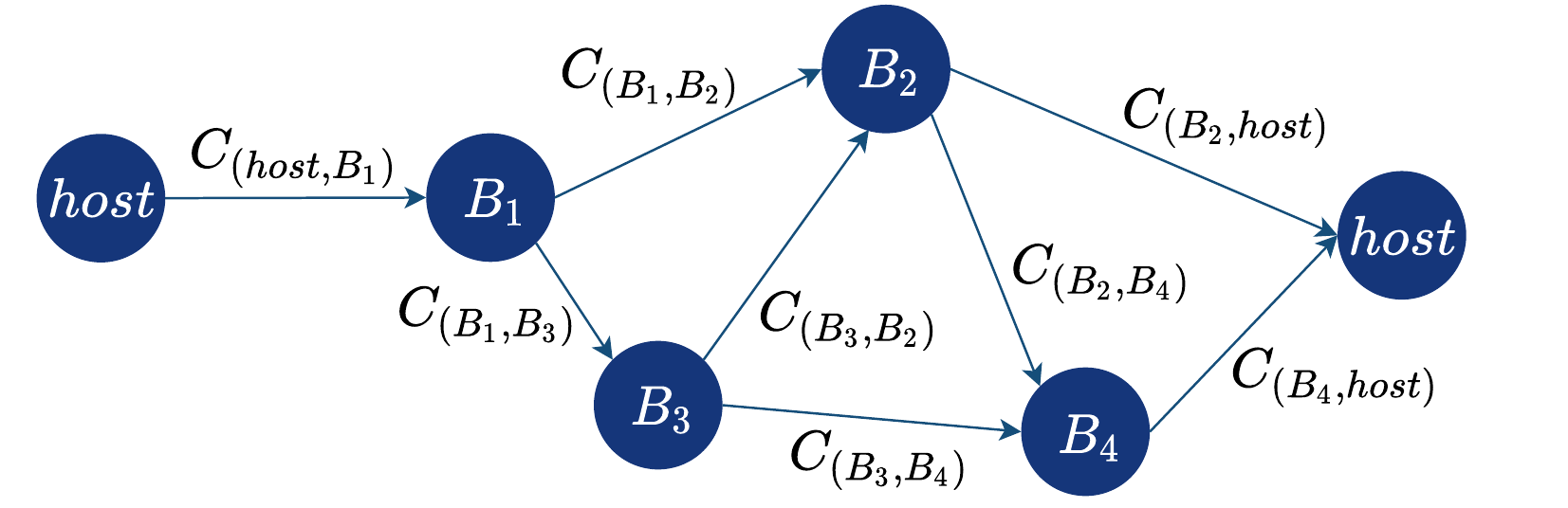}
    \caption{Graph representation of the interaction between the bacterial populations investigated in this paper, which are commonly found in the human GB.}
    \label{fig:FBA}
\end{figure}

The examples introduced in this section show that the human GB is a complex network to analyse. Therefore, we reduce the complexity of the human GB interaction model, even more, by considering the bacteria genera instead of the species. By taking this approach, we are able to represent the human GB with a few nodes and edges and, we exemplify this approach in Figure \ref{fig:FBA}. In this graph, we identified three types of interactions. First is the interactions between the host and the bacterial populations, where the bacterial populations consume nutrients (external inputs of the graph) from the host. These interactions are denoted as $C_{(host, B_k)}$, where the interaction starts with the host and ends with the bacterial population $B_k$, which acts as an inward node of the graph. The second type of interaction is identified when the host consumes the metabolite produced by the bacterial populations. These interactions are presented as $C_{(B_{k}, host)}$ where the interactions start with $B_k$ and end with the host where the bacterial population, $B_k$ acts as an outward node of the graph. Finally, when the metabolites produced by one bacterial population and consumed by other bacterial populations in the human GB, we classify them as direct interactions, $C_{(B_k, B_{k'})}$ where $B_{k'}$ is any bacterial population except $B_k$. Here, the bacterial populations $B_k$ and $B_k'$ act as relay nodes of the graph.

The bacterial populations in the human GB are capable of receiving and producing multiple metabolites that are processed through multiple pathways. Therefore, we can safely assume that all bacterial populations in the human GB consume or produce metabolites. Hence, the inward degree and outward degree of a node in our bacterial population graph layer must be $\text{deg}^-(B_{k}), \text{deg}^+(B_{k}) \geq 1$. 
We would also like to highlight that the outputs of this metabolism, metabolite production and the bacterial populations' growth play significant roles in the performance and the robustness of the human GB. Therefore, to analyse the contribution and role of each node in the graph, we consider the metabolism output of each bacterial population as the weight of the node. The metabolism output is modeled as the signal processing performance $SPP(B_{k})$, which is further detailed in Section \ref{sec:bottom_layer}.

\subsection{Bottom Layer - MC System}\label{sec:bottom_layer}

As detailed in the previous section, the metabolism of nutrients by the human GB involves the reception, processing, production of  metabolites. These activities are fundamental for the maintenance of the human GB, and this is modeled as the MC layer shown in Figure \ref{fig:TwoLayerModel}. Our aim of having the two-layer model is to determine how the changes due to molecular signals of the metabolites will affect the relationship of the bacterial population graph layer. Therefore, any changes in the bottom layer directly affect the upper layer and vice-versa.

Here, we model the bottom layer based on the definitions of the upper layer. Therefore, we define the metabolites as the molecular signals that are exchanged by the nodes, which can assume different functions depending on the MC network structure. For example, when the node receives molecular signals, we model it as a receiver, and when processing and secreting molecular signals we define them as transmitters based on the MC paradigm. The edges of the proposed MC network are represented as the MC channels to model the physical transport of molecular signals between the nodes. Figure \ref{fig:TwoLayerModel} shows a visual representation of the proposed bottom layer and its relationship with the upper layer.  

One of the functions that the nodes can execute is the molecular signal transmission and in this case, the emission of metabolites. Once the molecular signal is released to the extracellular space, it is diffused through the fluidic MC channel interconnecting the nodes. This process is governed by Fick's Law of diffusion and is represented as follows
 \begin{equation} 
\frac{\partial c(x,t)}{\partial t}=-D(x,t)\frac{\partial^2 c(x,t)}{\partial x^2},
\end{equation}
where $c(x,t)$ is the molecular signal concentration, $D$ is the diffusion coefficient, $x$ is the particle position (in metres) and $t$ is the time (in seconds).

The diffused molecular signal is received by the nodes which have the membrane receptors that will allow the metabolites  to bind. The performance of this network node function (i.e., molecular reception) relies on many factors such as  molecule size \cite{farsad2016comprehensive, chahibi2014molecular}, ligand-receptor maximum attraction length \cite{chahibi2014molecular}, binding noise due to the Brownian motion of molecules near the receptors \cite{kuscu2019transmitter}, ligand-receptor bond equilibrium \cite{chahibi2014molecular}, and the minimum required concentration to be detected \cite{llatser2013detection}. After receiving the molecular signals, the node will process them internally which may result in the production of a new molecular signal to be transmitted to the next node (focus of this paper) or consume the molecular signal. 

Upon reception of the molecular signals, the nodes execute a series of actions to internally processes them. These actions are often modelled as signalling pathways and the end result of this process are the metabolites that will be transmitted to the next node or to the host \cite{oliphant2019macronutrient}. 
\sm{The signalling process performance of a bacterial cell $b^{B_{k}}$ related to any metabolite $M_j$ is represented as $SPP_{b^{B_{k}}}(M_j)$. Let's assume that the cell $b^{B_{k}}$ produces $M_j$ by consuming another metabolite $M_{j'}$. 
Then the signal process performance $SPP_{b^{B_{k}}}(M_j)$ can be modeled by considering the metabolite $M_{j'}$ reception process (defined as $R_{b^{B_{k}}}(M_{j'})$), the encoding/decoding process from metabolite $M_{j'}$ to $M_j$ (defined as $E_{b^{B_{k}}}(M_{j'},M_{j})$), and $M_j$ metabolite secretion process by the cells in the bacterial population $b^{B_{k}}$ (defined as $S_{b^{B_{k}}}(M_{j})$). Hence, we represent the signal process performance as follows,}

\begin{equation}
    \label{eq:SPPCalc}
    SPP_{b^{B_{k}}}(M_{j})=f(R_{b^{B_{k}}}(M_{j'}),E_{b^{B_{k}}}(M_{j'},M_{j}),S_{b^{B_{k}}}(M_{j})).
\end{equation} 
Therefore, the SPP of the populations $B_{k}$ can be modeled as follows,
\begin{equation}
    \label{eq:NodeSPP}
    SPP_{B_{k}}(M_{j}) = \sum SPP_{b^{B_{k}}}(M_{j}).
\end{equation}
Since, the output of the molecular signal processing is the emission of a particular molecular signal, it is fair to say,
\begin{equation}
    \label{eq:SPPvsInteractions}
    SPP_{B_{k}}(M_{j}) = C^r_{(B_k, \Omega)}(M_{j})
\end{equation}
where $C^r_{(B_k, \Omega)}(M_{j})$ is the rate of molecule $M_j$ production by the bacterial population $B_k$ to any node (either other bacterial populations or host cells).

\begin{figure}[t!]
    \centering
    \includegraphics[width=\columnwidth]{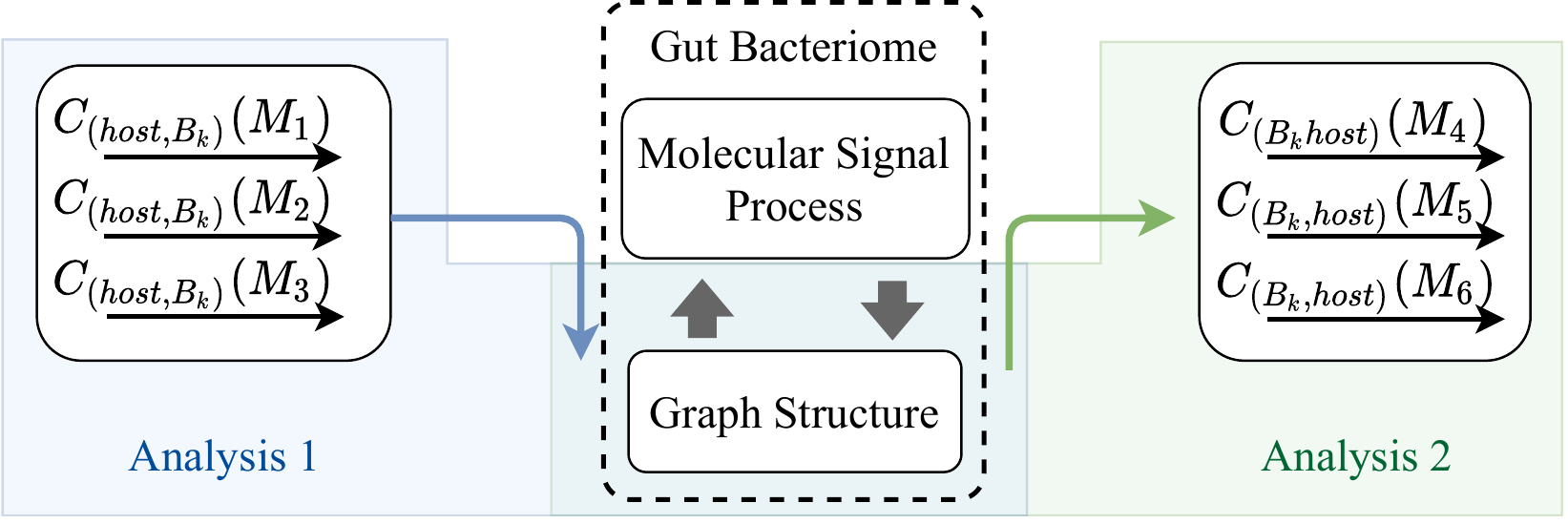}
    \caption{Illustration of the analysis structure of this study, where Analysis 1 focuses on the influence of inputs on the graph structure while Analysis 2 examine the behaviors of graph output against the structural deviations. See Sections \ref{sec: InputVsStructure} and \ref{sec:StructureVsSNR} for our analysis methods, and Sections \ref{sec:analysis1} and \ref{sec:StructureVSOutput} for our results.}
    \label{fig:analyses_structure}
\end{figure}

\section{System Dynamics}\label{sec:SystemDynamics}

Based on the proposed \sm{two-layer human GB model}, we investigate the system dynamics of the human GB through a series of simulations using a virtual GB model. \sm{To that end, we first recreate the digital form of human GB on the simulator. This recreation process is explained in depth later in Section \ref{sec:VirtualGB}.} Then we perform two main sets of experiments as depicted in Figure \ref{fig:analyses_structure}. In the first set, we analyse the impact of the system's inputs on the connectivity structure of the virtual GB as we understand that the molecular input signals may work as an altering factor of the human GB interaction patterns. In the second set, we manipulate the composition of our virtual GB to investigate the impact on the metabolite production that leaves the outward node of our MC network. Through this second set of experiments, we aim to identify the nodes that have the largest impact on the overall production of the metabolites and can play a pivoting role in the GB imbalances.

In our analyses, first we define a standard graph state $S_{0}$, which represents the functionality of an average healthy human GB with the intention of quantifying structural connectivity changes and behavioral deviations that is relative to the standard structures. The average composition, interactions and metabolite production dynamics were mainly considered in defining the $S_{0}$. Finally, the accuracy of the $S_{0}$ is confirmed using the output metabolite ratios over the exact values as the virtual GB scales down the ecosystem. The average composition and the interactions of $S_{0}$ for the case study of this paper is presented in Section \ref{sec:AnalyticalResults}

\subsection{Molecular input impact on the human GB structure}\label{sec: InputVsStructure}

Due to the variety of bacterial behaviors induced by the exchange of molecules, some of the molecular input signals have a significant impact on the structure of the human GB (our focus), while others are directly converted into output metabolites. In this section, we detail how the molecular input signals impact the structure of our MC network. The structural deviations of the graph is a crucial measurement in understanding the deviation of the human GB behavior from the healthy state. This structural deviation is evaluated in terms of edges and nodes weight variations using the rates of the interaction of the nodes. Hence, we explain how the interaction rates can be calculated theoretically using FBA and are represented as follows

\begin{equation}\label{eq:FBA1}
    F_{[k\times q]} \cdot \vec{C} = \vec{M}_{Con}(B_{k})
\end{equation}
where $F_{[k\times q]}$ is the stoichiometric matrix of $k$ number of bacterial populations and $q$ number of interactions based on the flux of metabolites between the nodes in the MC network.
Here $\vec{C} =[C^r_{1}, C^r_{2},..., C^r_{q}]_{1\times q}$ and $C^r_q$ is the rate of  interactions for $C_q$. We can solve (\ref{eq:FBA1}) as follows,
\begin{equation}\label{eq:int_rate}
\begin{blockarray}{ccccc}
\begin{block}{c(cccc)}
 B_{1} & a_{1,1}  & a_{1,2}  &... & a_{1,q}\\ 
 B_{2} & a_{2,1}  & a_{2,2}  &... & a_{2,q} \\ 
 \vdots  &\vdots  &\vdots    &\ddots  &\vdots \\ 
 B_{k} & a_{k,1}  & a_{k,2}  &... &a_{k,q}\\
\end{block}
\end{blockarray}
\cdot
\begin{blockarray}{c}
\begin{block}{(c)}
 C^r_{1}\\ 
 C^r_{2}\\ 
\vdots \\
 C^r_{q}\\
\end{block}
\end{blockarray}
=
\begin{blockarray}{c}
\begin{block}{(c)}
\frac{\mathrm{d}M_{Con}(B_{1})}{\mathrm{d}t}\\ 
\frac{\mathrm{d}M_{Con}(B_{2})}{\mathrm{d}t}\\ 
\vdots \\
\frac{\mathrm{d}M_{Con}(B_{k})}{\mathrm{d}t}\\
\end{block}
\end{blockarray}
\end{equation}
where, $a_{k,q}$ is the stoichiometry of the interaction $C^r_q$ for bacterial population $B_k$. 

\sm{Based on (\ref{eq:int_rate})}, we can extract the relationship between rates of interactions starting from the node $B_{k}$ using Mass Balance Equation (MBE) as, which is based on the following relationship
\begin{equation}
\frac{\mathrm{d}M_{Con}(B_{k})}{\mathrm{d}t}=\sum_{q}a_{(k,q)}C^r_{q}.
\end{equation}
On the other hand, the rate of molecular consumption can be modeled as follows \cite{martins2016using}, 
\begin{equation}
    \frac{\mathrm{d}M_{Con}(B_{k})}{\mathrm{d} t}=-U_{1}\left ( \mu_{k}\frac{M_{Con}(B_{k})}{M_{Con}(B_{k})+K_{S1}} \right )N_{B_{k}}
\end{equation}
where $N_{B_{k}}$ is the bacterial concentration, $\mu_{k}$ is maximum growth rate, $K_{S1}$ is the half-saturation constant of the bacteria, and $U_{1}$ is an utility parameter.
Hence, 
\begin{equation}
    -U_{1}\left ( \mu_{k}\frac{M_{Con}(B_{k})}{M_{Con}(B_{k})+K_{S1}} \right )N_{B_{k}}=\sum_{q}a_{(k,q)}C^r_{q}.
\end{equation}
By solving the series of MBEs, all the interaction rates can be calculated. This is a highly complex calculation due to the massive number of nodes and the edges of the network. Additionally, there is a large number of parameters associated to the structural connections that influence the MBEs. In order to minimize this complexity, the virtual GB produces data that is related to the rates of interactions, and this will avoid the complex FBA calculations. 

The extracted rates of interactions are then used to quantify the graph structural changes in two ways. First, we investigate the graph structural changes considering the behaviors of the node weights. Here, the statistical distances between the weights of the same node in different graph states are measured. The node weights of this system are modeled using the $SPP$. Collective metabolite production of the bacterial population $B_{k}$ in the graph state $S_g$ is considered as the node weight $B_{k}^{w}(S_g)$ and can be evaluated as follows,
\begin{equation}\label{eq:node_weight}
    B_{k}^{w}(S_g)=\sum_{j}SPP_{B_{k}}(M_{j}).
\end{equation}
Alternatively, using (\ref{eq:SPPvsInteractions}) we compute the node weight as follows,
\begin{equation}\label{eq:node_weightbyInteraction}
    B_{k}^{w}(S_g)=\sum_{j}C^r_{(B_k,\Omega)}(M_{j}).
\end{equation}
Based on this, $d(B^w_{k}:S_{g},S_{0})$ that represents the distance of node $B_{k}$ between the two graph states $S_{0}$ and $S_{q}$ is evaluated as follows,
\begin{equation}
\label{eq:NodeHammingDistance}
    d(B^w_{k}:S_{g},S_{0}) = B_{k}^{w}(S_g)-B_{k}^{w}(S_0).
\end{equation}

Next, we quantify the structural deviation of the graph using the interaction changes.
There are several methods from the literature for distance calculation between two graph states. In this study, we consider static snapshots of different graph states that can enable the use of the \emph{Hamming Distance} to evaluate graphical distances for two states \cite{deza2009encyclopedia}. The Hamming distance $d_{h}(S_{0}, S_{g})$ between the graph states $S_g$ and the standard state $S_0$ is defined as the difference of two adjacent matrices corresponding to the two graph states. First, we define the adjacency matrix of the graph state $S_{g}$ as follows,

\begin{equation}
\begin{blockarray}{ccccc}
 & B_{1} & B_{2} & ... & B_{k}\\
\begin{block}{c(cccc)}
 B_{1} & C^w_{(B_{1},B_{1})}  & C^w_{(B_{1},B_{2})}  &... & C^w_{(B_{1},B_{k})}\\ 
 B_{2} & C^w_{(B_{2},B_{1})}  & C^w_{(B_{2},B_{2})}  &... & C^w_{(B_{2},B_{k})} \\ 
 \vdots  &\vdots  &\vdots    &\ddots  &\vdots \\ 
 B_{k} & C^w_{(B_{k},B_{1})}  & C^w_{(B_{k},B_{2})}  &... & C^w_{(B_{k},B_{k})}\\
\end{block}
\end{blockarray}
\end{equation}
where $C^w_{(B_{k},B_{k})}$ is the weight of the interaction $C_{(B_{k},B_{k})}$. 

\begin{figure}[t!]
    \centering
    \includegraphics[width=\columnwidth]{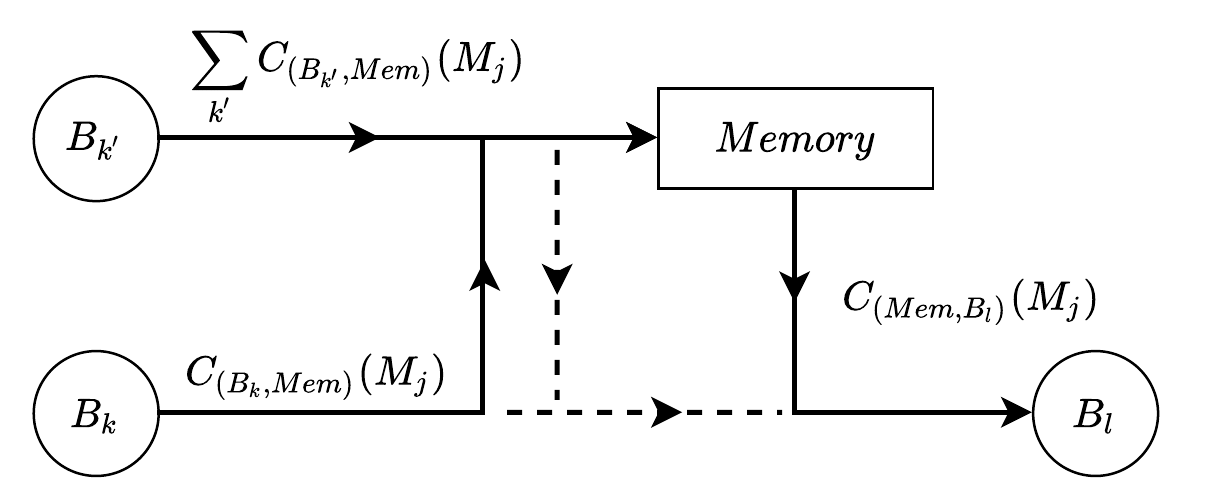}
    \caption{Illustration of the environment working as a memory of molecules.}
    \label{fig:MemModule}
\end{figure}

\sm{We define the weight of the interaction $C_{(B_k,B_{k'})}(M_j)$ as follows,}
\begin{equation}
\label{eq:EdgeWeighEmpirical}
    C^w_{(B_k,B_{k'})}(M_j)=\frac{C^r_{(B_k, \Omega)}(M_j)}{\sum_{k}C^r_{(B_k, \Omega)}(M_j)}\cdot \frac{C^r_{(\Omega, B_k')}(M_j)}{\sum_{B_k'}C^r_{(\Omega, B_k')}(M_j)}.
\end{equation}
\sm{Moreover, from the number of molecules released by a bacterial population, only a small portion is consumed directly by the other populations and the remaining concentration will get accumulated in the environment.}
This means the most significant portion of molecular consumption by the bacterial populations is from the environment. We define this process with the help of a memory component concept as elaborated in Figure \ref{fig:MemModule}. \sm{Since the metabolites are accumulated in the environment, we consider it as memory, then model the metabolite accumulation as an interaction starting from a bacterial population that releases the metabolites and ending with the memory, $C_{(B_k, Mem)}$. In the same way, the metabolite consumption from the environment is modeled as an interaction starting from the memory and ending with a bacterial population that consumes the particular metabolite, $C_{(Mem, B_k)}$.}
Hence, we modify (\ref{eq:EdgeWeighEmpirical}) by applying the memory and is represented as follows,
\begin{equation}
\label{eq:EdgeWeighWithMem}
\begin{split}
    C^w_{(B_k,B_{k'})}(M_j)=\frac{C^r_{(B_k, Mem)}(M_j)}{\sum_{k}C^r_{(B_k, Mem)}(M_j)}\\[1mm]
    \cdot \frac{C^r_{(Mem, B_k')}(M_j)}{\sum_{B_k'}C^r_{(Mem, B_k')}(M_j)}.
\end{split}
\end{equation}

\sm{Then, the Hamming distance, $d_{h}(S_{0}, S_{g})$ can be represented as,
\begin{equation}
    d_{h}(S_{0}, S_{g})=\sum_{k,k'} |C_{(B_{k},B_{k'})}^{w}(S_g)- C_{(B_{k},B_{k'})}^{w}(S_0)|
\end{equation}
where, $C_{(B_{k},B_{k'})}^{w}(S_g)$ and $C_{(B_{k},B_{k'})}^{w}(S_0)$ are the weights of interaction $C_{(B_{k},B_{k'})}$ in graph states $(S_g)$ and $S_0$ respectively.}

\subsection{Human GB structure impact on the molecular output}
\label{sec:StructureVsSNR}

This analysis explores the impact of interaction variations of the human GB  on the output. Here, we keep the inputs at an optimal level and manually alter the graph structure by changing the population sizes which leads to variations in the $SPP$ of the nodes. Then the output of the system is measured in different graph states and the weights of the edges are calculated using (\ref{eq:EdgeWeighEmpirical}) to determine the molecular output of the MC layer using graph theory.

 \sm{The ratio between the three SCFAs can be identified as a critical measurement to evaluate the metabolite production accuracy of the bacteriome. We adopt SNR to evaluate the consistency of the output signal ratios. In this analysis, we calculate SNR of any signal $SNR(M_j)$, considering the other output signals, $M_{j'}$ as noise. This SNR value directly indicates the ratio between the molecular signal $M_j$ and other metabolite signals $M_{j'}$. Then $SNR(M_j)$ is calculated as follows,}

\begin{equation}
\label{eq:SNRCalulation}
SNR(M_j)=\displaystyle\sum_{k}\frac{C_{(B_k,host)}(M_{j})}{\sum_{j'} C_{(B_k,host)}(M_{j'})}.
\end{equation}



\sm{Moreover, some  bacterial populations do not produce specific SCFAs, but have an indirect influence on them. For example, \textit{Bacteroides} cells do not produce butyrate, but the acetate produced by the \textit{Bacteroides} cells is a substrate for the butyrate production by \textit{Faecalibacterium} and \textit{Roseburia} cells. Hence, the impact of changes in \textit{Bacteroides} population sizes cascades through the interaction network and affects the butyrate production. Considering the above mentioned effect, a correlation matrix is generated for variation of node weights vs the collective SCFA output of the human GB to analyse the impact of various bacterial populations in SCFA production. 
Here, we denote the rate of SCFA $M_j$ output by all the bacterial populations as $O^r(M_j)$ where}

\begin{equation}
    O^r(M_j) = \sum_{k} C^r_{(B_k, host)}(M_j).
\end{equation}
Then, the correlation coefficient $r(B_k)$ of node weight $B_k^w$ versus the collective output of $M_j$ is calculated based on the following relationship.


\begin{equation}
\begin{split}
r(B_k) =\displaystyle\sum_{g}\frac{\overline{B^w_{k}}-B^w_{k}(S_{g})}{\sqrt{(\overline{B^w_{k}}-B^w_{k}(S_{g}))^2}}\\[1mm]
\cdot\frac{(\overline{O^r}(M_j)-O^r(M_j))}{\sqrt{(\overline{O^r}(M_j)-O^r(M_j))^2}}
\end{split}
\end{equation}
where, $\overline{O^r}(M_j)$ is the standard collective output rate for $M_{j}$ by all the bacterial populations and $\overline{B^w_{k}}$ is the weight of the node $B_k$ in the standard state $S_0$.

\section{Analytical Results}\label{sec:AnalyticalResults}

In this section, we describe the development of the virtual GB and the results from our analysis that is based on the models presented in Section \ref{sec:SystemDynamics}. 

\subsection{Virtual GB Design}
\label{sec:VirtualGB}

We developed the virtual GB using the metagenomic data to characterize the bacterial populations signaling interactions and its impact on the network relationships. The virtual GB is inspired by the BSim agent-based cell simulator \cite{Gorochowski2012}. 
The virtual GB is written in C++ with CUDA platform for parallel processing to increase the simulation performance and most importantly, mimic the parallel processing typically executed by the bacterial  cells. \sm{We dedicate one GPU block for each bacterial cell, and the threads of that block to intracellular functions of the corresponding cell.} To simulate the bacterial interactions, we model the exchange of molecules using metabolic flux in a diffusive media. The simulator is equipped with a 3D environment, which can be divided into chambers to place host or bacterial cells in specific regions of the environment. The 3D environment is further designed using a voxel architecture as shown in Figure \ref{fig:VGB} which provides the ability of extracting data on metabolites interactions and accumulations separately. Moreover, we can introduce any new cell type by creating their internal metabolic pathways and other physiological characteristics such as motility, shape, size, etc. Therefore, the simulator can be used for a range of setups including other metabolic function, microbial ecosystems in different body habitats or targeting specific bacterial behavior like quorum sensing.

\sm{We designed this simulator to extract multi-dimensional data from the MC network of the human GB, and its data extraction capabilities can be divided into two categories.} In the case of data extraction from bacterial cells, the simulator can log data on the metabolite consumption, metabolite production and growth of every bacterial population. The simulator also generates data of the bacterial ecosystem environment. This includes data on consumption of nutrients, metabolite accumulation, and the locations of the bacterial cells and host cell can be easily determined. In this study, we setup the virtual GB to simulate the SCFA production using the metagenomic data obtained in \cite{oliphant2019macronutrient}, KEGG \cite{kanehisa2000a, kanehisa2019a, kanehisa2021a} and MetaCyc databases \cite{caspi-a} 
\begin{figure}[t!]
    \centering
    \includegraphics[width=\columnwidth]{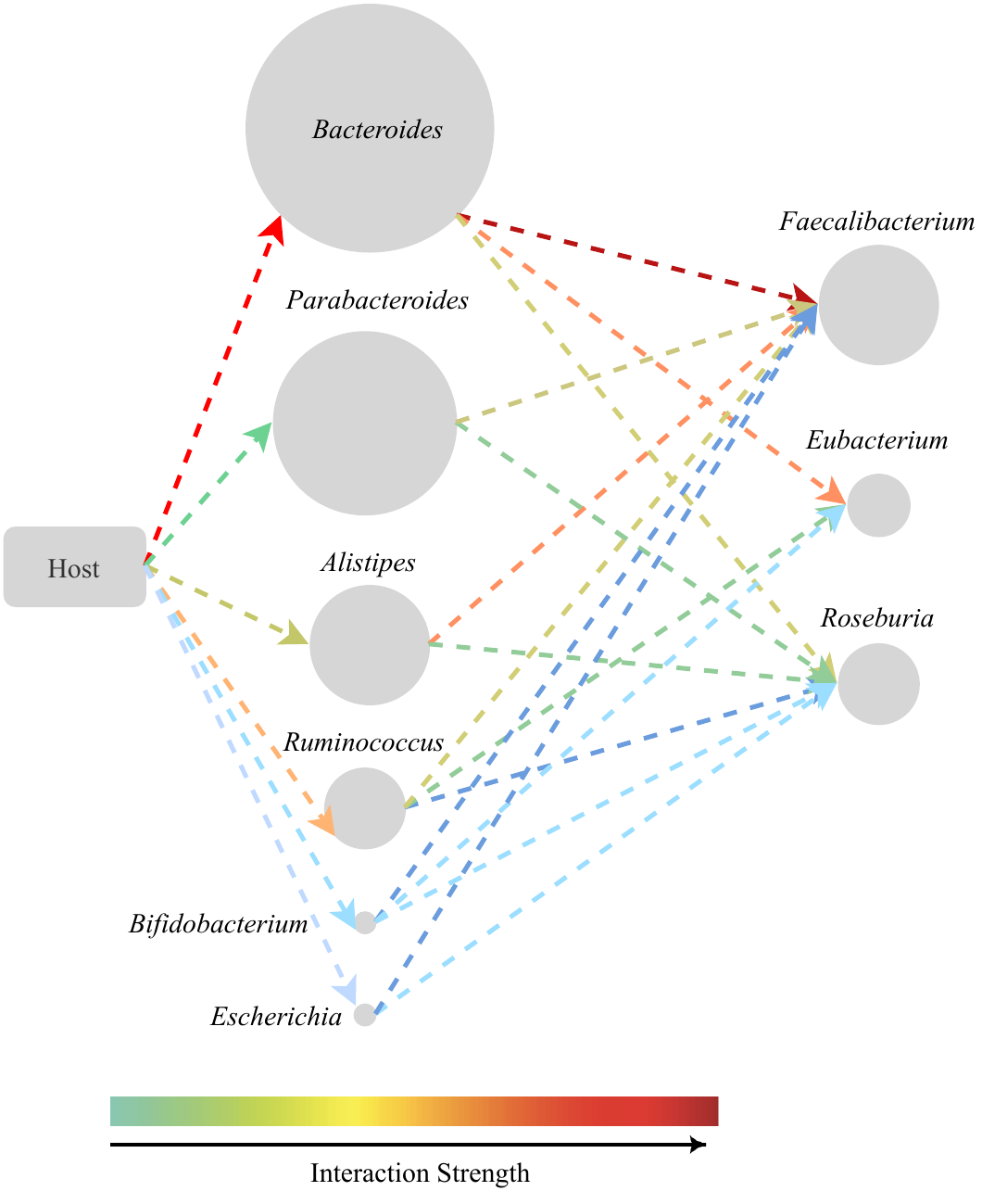}
    \caption{Representation of the subgraph considered in the case study which contains the nodes and edges related to SCFA production.}
    \label{fig:SCFAMetabolism}
\end{figure}
\begin{figure}[t!]
    \centering
    \includegraphics[width=\columnwidth]{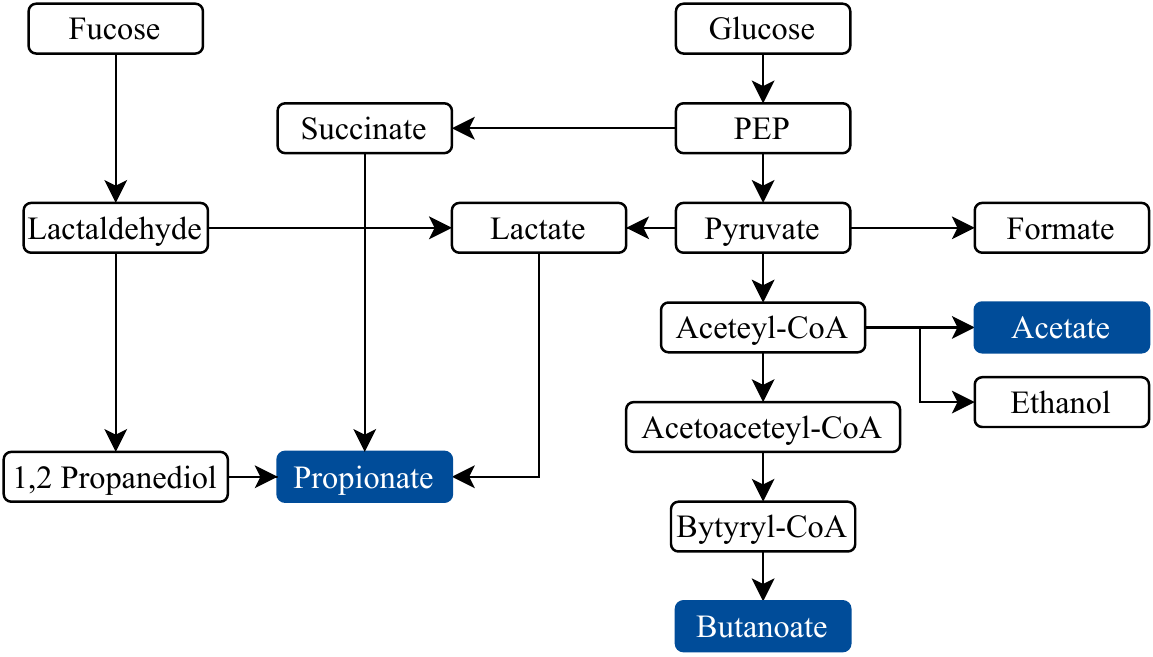}
    \caption{Combined and simplified SCFA production pathway of converting fucose and glucose, into SCFAs.}
    \label{fig:SCFAProduction}
\end{figure}
\begin{figure*}[t!]
    \centering
    \includegraphics[width=\textwidth]{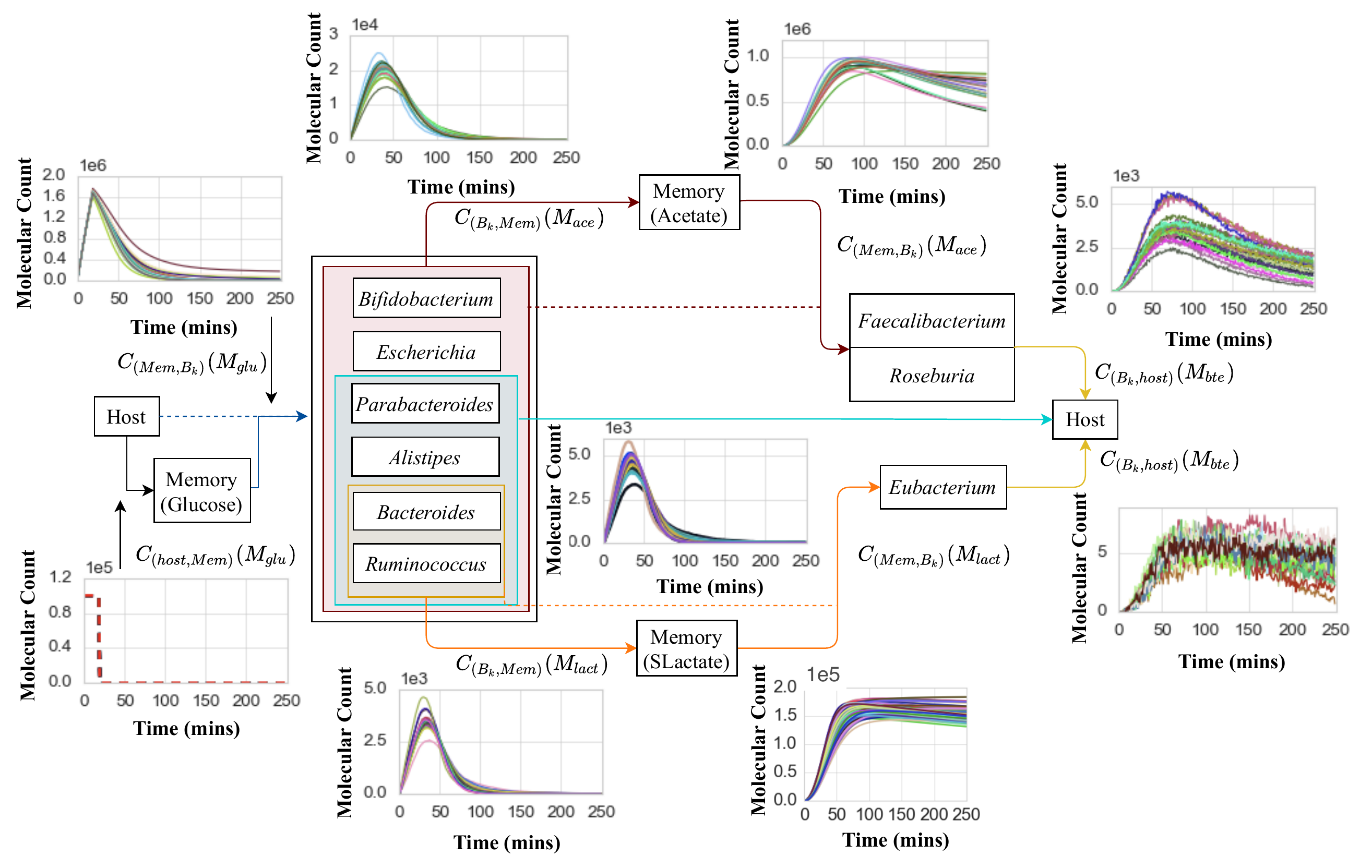}
    \caption{Representation of the simplified MC network with plots illustrating link behaviors for different compositional changes. Please note that $glu$, $ace$, $lact$ and $bte$ stands for glucose, acetate, lactate and butyrate respectively.}
    \label{fig:A2-Network_SimplifiedWithGraphs}
\end{figure*}
Here, we present a series of analyses conducted on the SCFA production within the human GB using the two-layer model of the Bacterial Population Graph layer and the Molecular Communication Layer. SCFA are the main products of fermentation of non-digestible carbohydrates\cite{morrison2016formation}. First, we defined the average composition of the human GB using the relative abundance (RA) data extracted from 352 samples of the MicrobiomeDB \cite{MicrobiomeDB}. \sm{In-depth graphical representation of RA data of all these samples} are shown in the Appendix. Then, we calculate the average RAs for the most abundant genera related to the SCFA production, which is shown in Table \ref{table:RA}.

\begin{figure*}[t!]
\centering
\subfloat[\label{fig:A1-GlucoseSubGraph}]{
\includegraphics[width=0.28\textwidth]{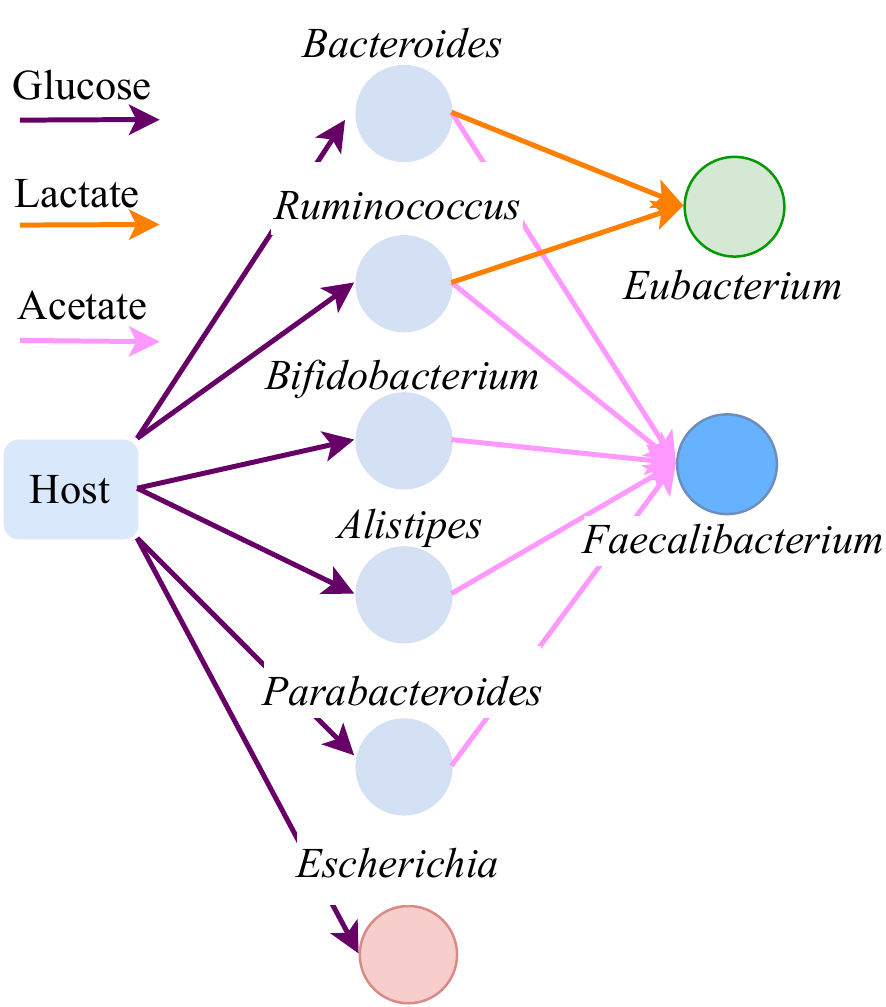}}
\subfloat[\label{fig:A1-GlucoseMiddleLayer.pdf}]{
\includegraphics[trim=10 14 10 0,clip,width=0.36\textwidth]{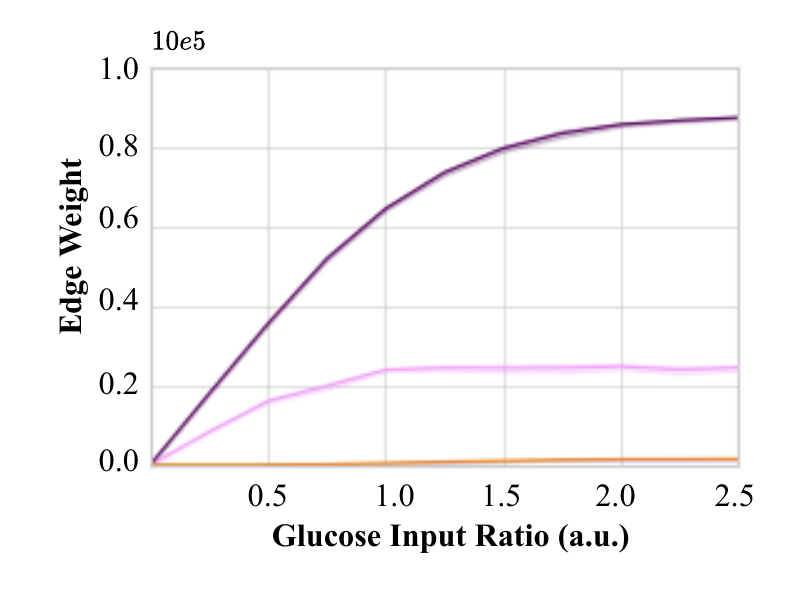}}
\subfloat[\label{fig:A1-GlucoseMCLayer.pdf}]{
\includegraphics[trim=10 14 10 0,clip,width=0.35\textwidth]{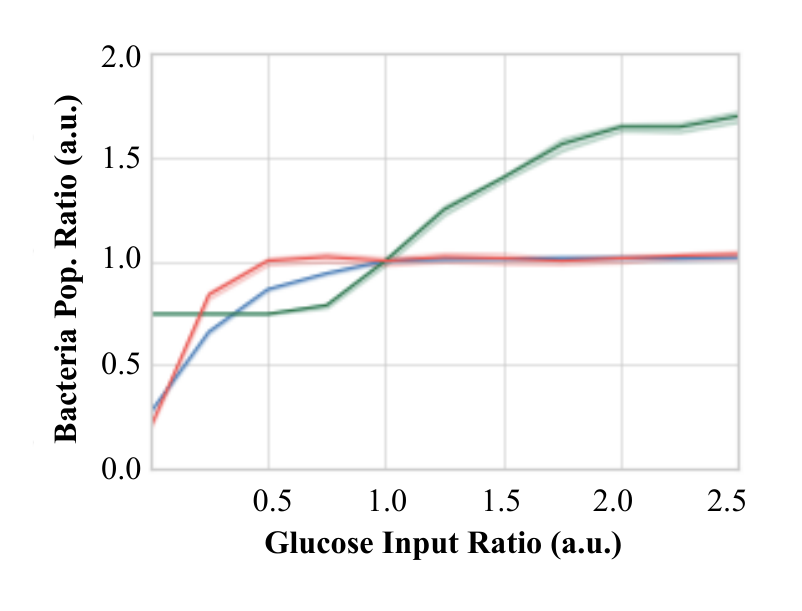}}
\caption{Deviation of population sizes of \textit{Faecalibacterium}, \textit{Eubacterium} and Escherichia from the optimal levels due to different input concentrations of glucose: (a) subgraph for the glucose consumption, (b) edge weight behaviors of the intermediate interactions, and (c) population growth behaviors.}
\label{fig:A1-GlucoseVsGrowth}
\end{figure*}

\begin{figure*}[t!]
\centering
\subfloat[\label{fig:A2-BacteroidesSubGraph}]{
\includegraphics[width=0.32\textwidth]{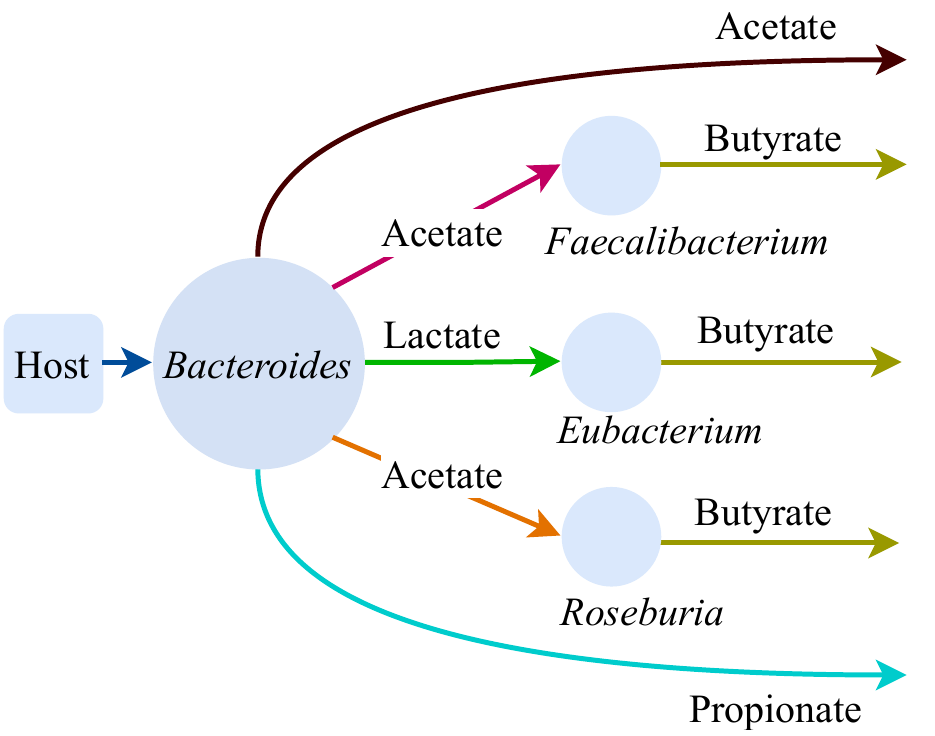}}
\subfloat[\label{fig:A2-BacteroidesMiddleLayer}]{
\includegraphics[trim=0 10 0 0,clip,width=0.33\textwidth]{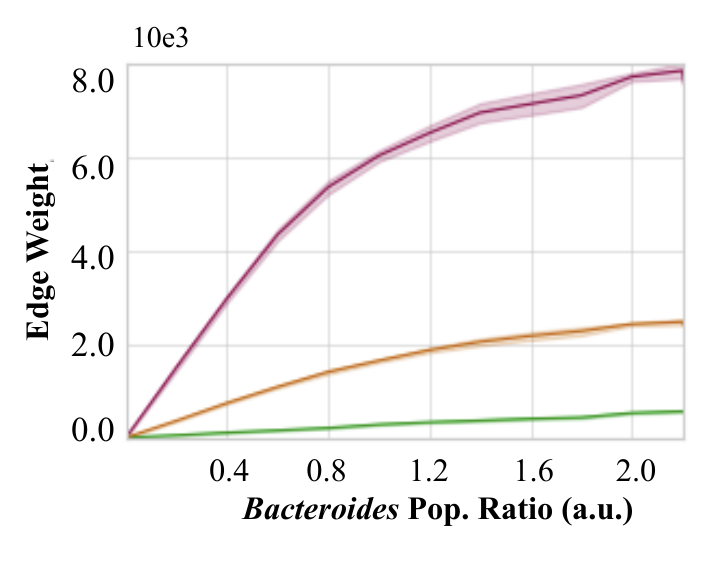}}
\subfloat[SCFA Output\label{fig:A2-BacteroidesMCLayer}]{
\includegraphics[trim=0 10 0 0,clip,width=0.34\textwidth]{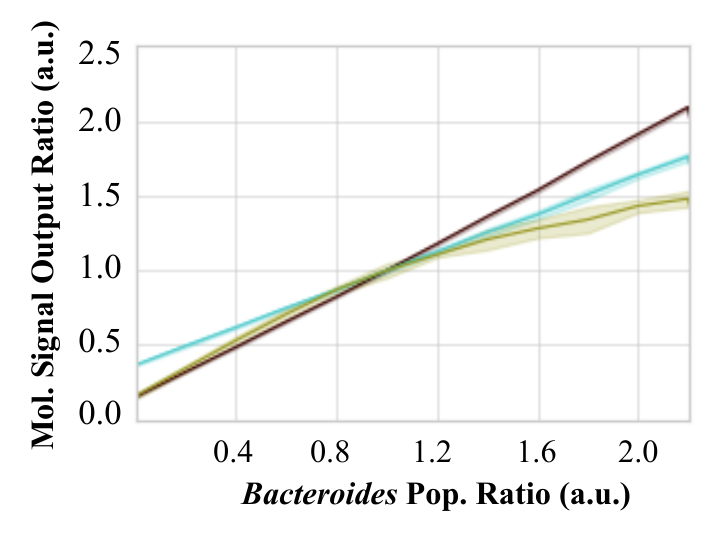}}
\caption{Behaviors of SCFA production for various in \textit{Bacteroides} population sizes: (a) subgraph of Bacteroides population interactions, (b) edge weight behaviors, and (c) SCFA output.}
\label{fig:A2-BacteroideVsSCFA}
\end{figure*}

\begin{figure*}[t!]
\centering
\subfloat[\label{fig:A2-FaecalibacteriumSubGraph}]{
\includegraphics[width=0.34\textwidth]{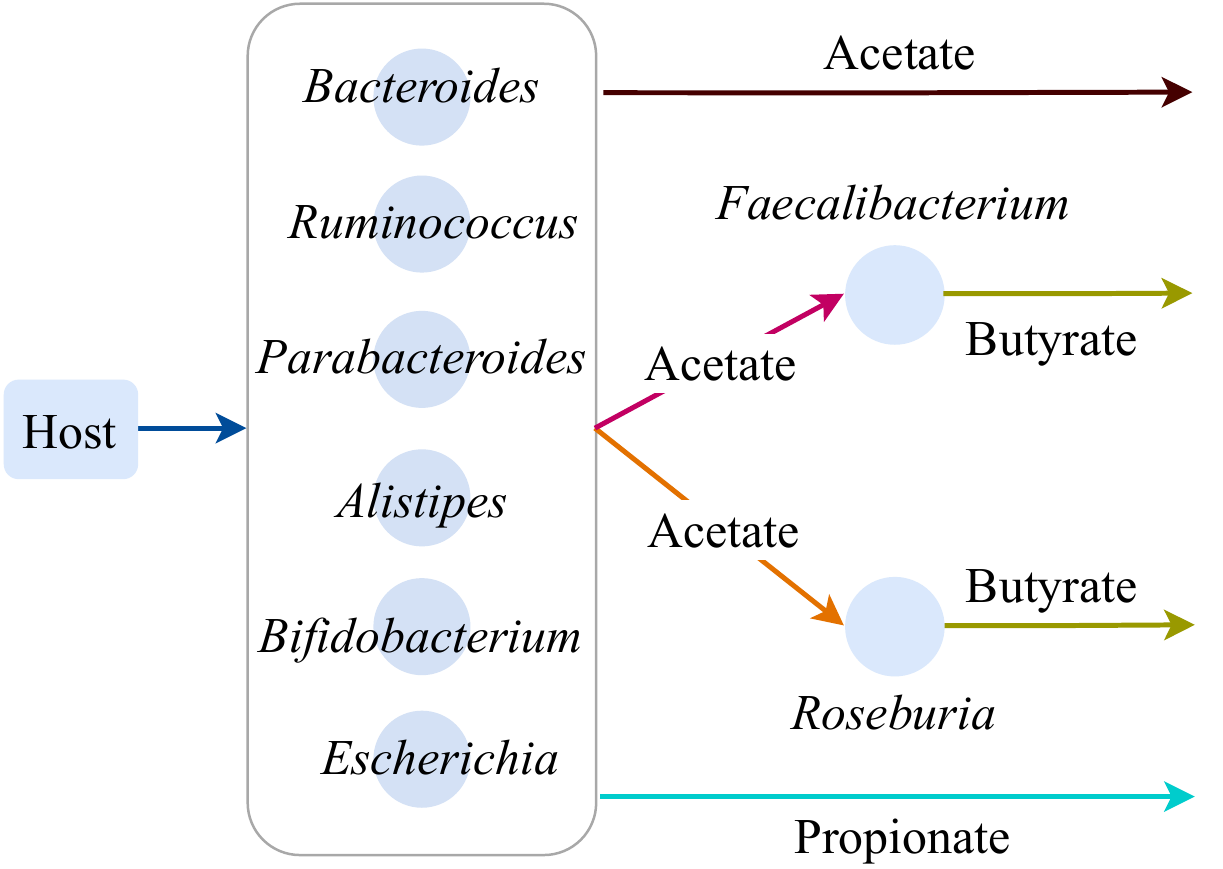}}
\subfloat[\label{fig:A2-FaecalibacteriumMiddleLaye}]{
\includegraphics[trim=5 10 0 0,clip,width=0.33\textwidth]{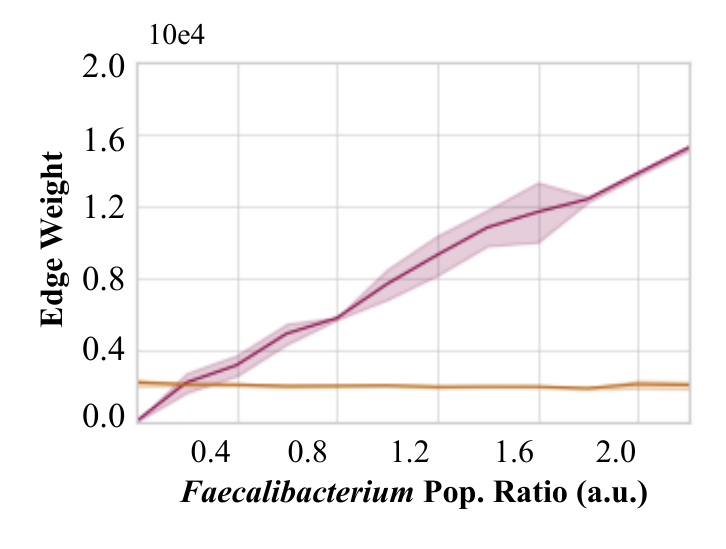}}
\subfloat[\label{fig:A2-FaecalibacteriumMCLayer}]{
\includegraphics[trim=5 10 0 0,clip,width=0.32\textwidth]{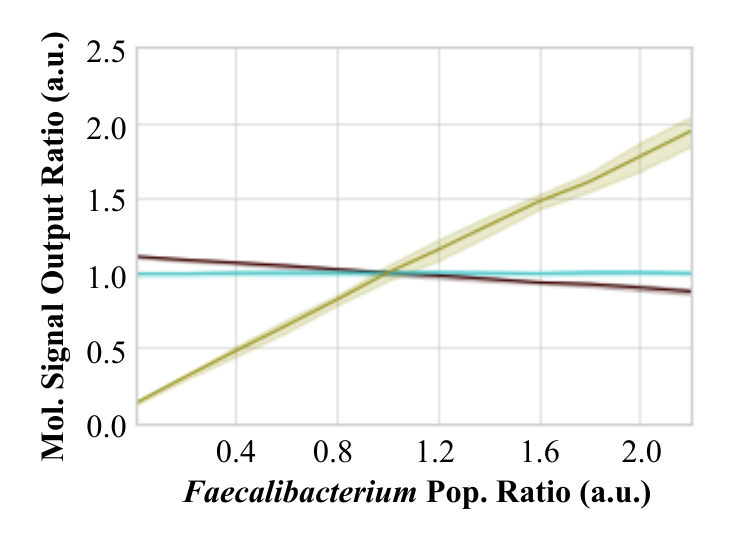}}
\caption{Behaviors of SCFA production for various in \textit{Faecalibacterium} population sizes: (a) subgraph related to the interactions of \textit{Faecalibacterium} population, (b) edge weight behaviors, and (c) SCFA output.}
\label{fig:A2-FaecaliVsSCFA}
\end{figure*}

\begin{figure}
\centering
\begin{subfigure}[b]{\linewidth}
\includegraphics[width=\linewidth]{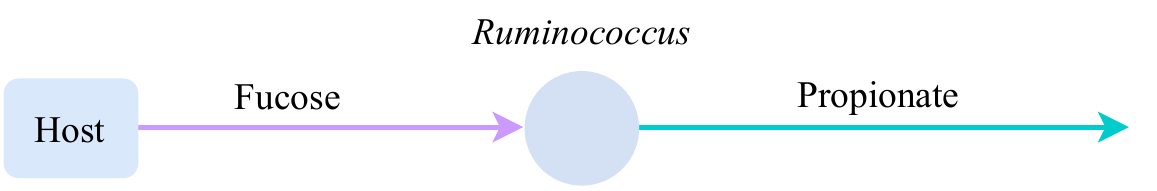}
\caption{}\label{fig:A2-RuminococcusSubGraph}
\end{subfigure}

\begin{subfigure}[b]{.49\linewidth}
\includegraphics[width=\linewidth]{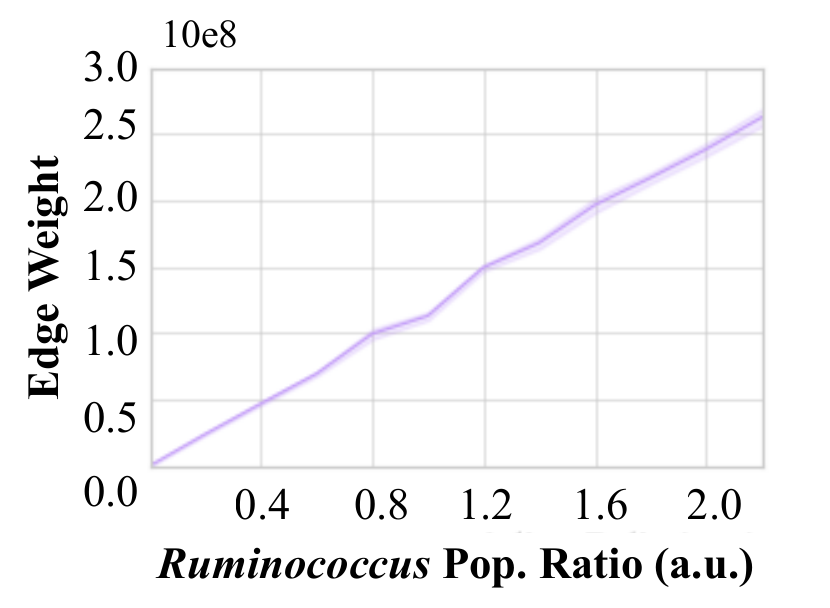}
\caption{}\label{fig:A2-RuminococcusMiddleLayer}
\end{subfigure}
\begin{subfigure}[b]{.49\linewidth}
\includegraphics[width=\linewidth]{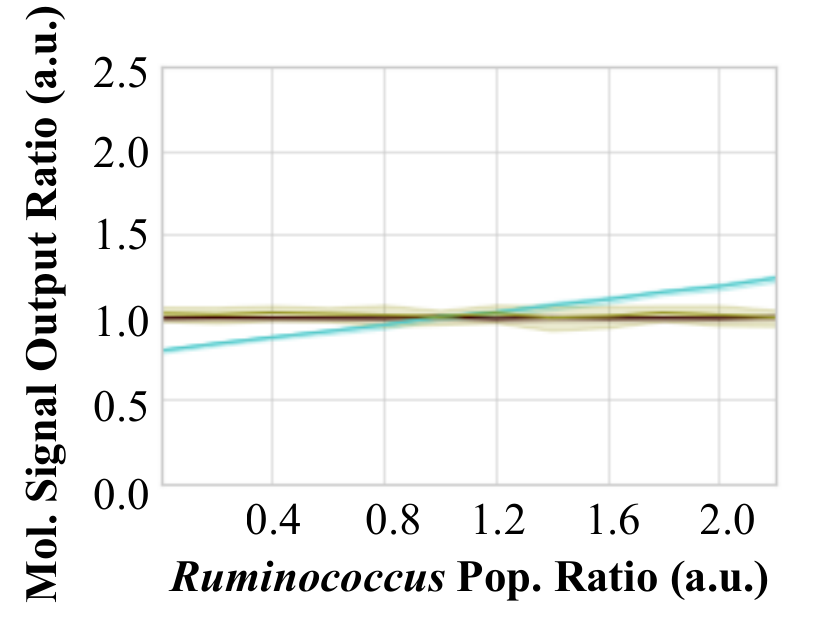}
\caption{}\label{fig:A2-RuminococcusMCLayer}
\end{subfigure}

\caption{Behaviors of SCFA production for various in \textit{Ruminococcus} population sizes: (a) subgraph related to the interactions of \textit{Ruminococcus} population, (b) edge weight behavior, and (c) SCFA output.}
\label{fig:A2-RuminococcusVSSCFA}
\end{figure}

\sm{Using these RA data along with the extracted interaction data from the databases mentioned earlier,} 
 we created a graph network for SCFA production, $\Gamma_{SCFA}$ \sm{following the definitions presented in Section \ref{sec:upper_layer}}.
Figure \ref{fig:SCFAMetabolism} illustrates the $\Gamma_{SCFA}$ where nodes sizes of Figure are proportional to the RAs of the respective genera shown in Table \ref{table:RA}. Furthermore, the edges are color coded to highlight the strengths of the interactions which are quantified using  (\ref{eq:EdgeWeighEmpirical}). The direct interactions between the host and the first layer nodes represent the GB consumption of glucose and fucose supplied by the host. The interaction from the host to the \textit{Bacteroides} cells through glucose, $C_{(host, Bact)}(M_{glu})$, and \textit{Bacteroides} cells to \textit{Faecalibacterium} cells through acetate, $C_{(Bact, Fae)}(M_{ace})$, are the strongest since the \textit{Bacteroides} cells dominate the metabolism process. The interactions from the first layer of bacterial population nodes to the second layer are composed of interactions through lactate as well as acetate, which are shown later in detail in Figures \ref{fig:A1-GlucoseSubGraph}, \ref{fig:A2-BacteroidesSubGraph}, \ref{fig:A2-FaecalibacteriumSubGraph} and \ref{fig:A2-RuminococcusSubGraph}.

\sm{
For illustration purposes, we combine the metabolic processes executed on different bacterial cells and simplify the SCFA pathway to focus on the most important steps that leads to the production of the three most abundant SCFAs in the human GB, namely acetate, butyrate and propionate (see Figure \ref{fig:SCFAProduction})\cite{rowland2018gut}. Please note that in a typical human GB their abundance ratios range from 3:1:1 to 10:2:1  \cite{den2013role}, and we utilize this metric to establish the normal behavior of the bacterial ecosystem within our virtual GB. This enables us to modify the MC network inside the virtual GB to better investigate the effects of the molecular interactions between nodes. An example of such investigation can be seen in Figure \ref{fig:A2-Network_SimplifiedWithGraphs}, where each link is an interaction between the graph nodes in our virtual GB. Furthermore, each plot in Figure \ref{fig:A2-Network_SimplifiedWithGraphs} shows the behavior of the molecular signal transported in each link and their  multiple molecular count values result from the different bacterial population sizes considered for each link. More detailed results for the virtual GB compositional changes are explained in following sections. Please note that the high variation in the molecular count shown in the interactions between \textit{Faecalibacterium}/\textit{Roseburia}, \textit{Eubacterium} and host cells (see Figure \ref{fig:A2-Network_SimplifiedWithGraphs}) are due to diffusion effects and are not treated as noise in this work. Therefore, our results are based on the average MC network behavior after twenty runs of the virtual GB to take this effect into account.}

\begin{table}[t!]
\centering
\caption{Average RAs of bacterial populations}
\label{table:RA}
\begin{tabular}{l|l}
\hline
Genus            & Average RA \\ \hline
\textit{Bacteroides}      & 0.4899173  \\
\textit{Alistipes}        & 0.05960802 \\
\textit{Faecalibacterium} & 0.04329791 \\
\textit{Parabacteroides}  & 0.04096428 \\
\textit{Ruminococcus}     & 0.03320183 \\
\textit{Roseburia}        & 0.01039938 \\
\textit{Eubacterium}      & 0.0093219  \\
\textit{Bifidobacterium}  & 0.00179366 \\
\textit{Escherichia}      & 0.00185639 \\ 
\hline
\end{tabular}
\end{table}

\subsection{Analysis 1: Molecular input effects on the graph structure}\label{sec:analysis1}

\begin{figure}[t!]
    \centering
    \includegraphics[trim=5 0 0 0,clip,width=\columnwidth]{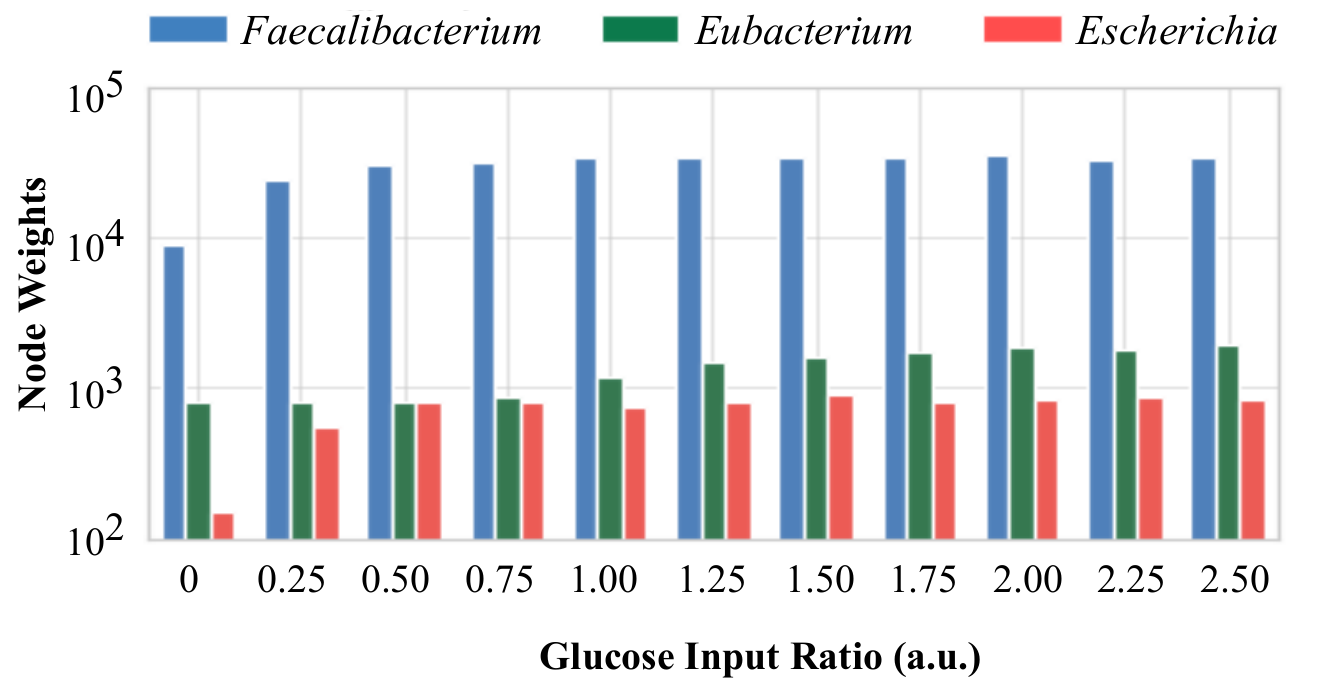}
    \caption{Changes of node weights due to the variations in molecular signal inputs.}
    \label{fig:Weights}
\end{figure}

\begin{figure}[t!]
    \centering
    \includegraphics[width=\columnwidth]{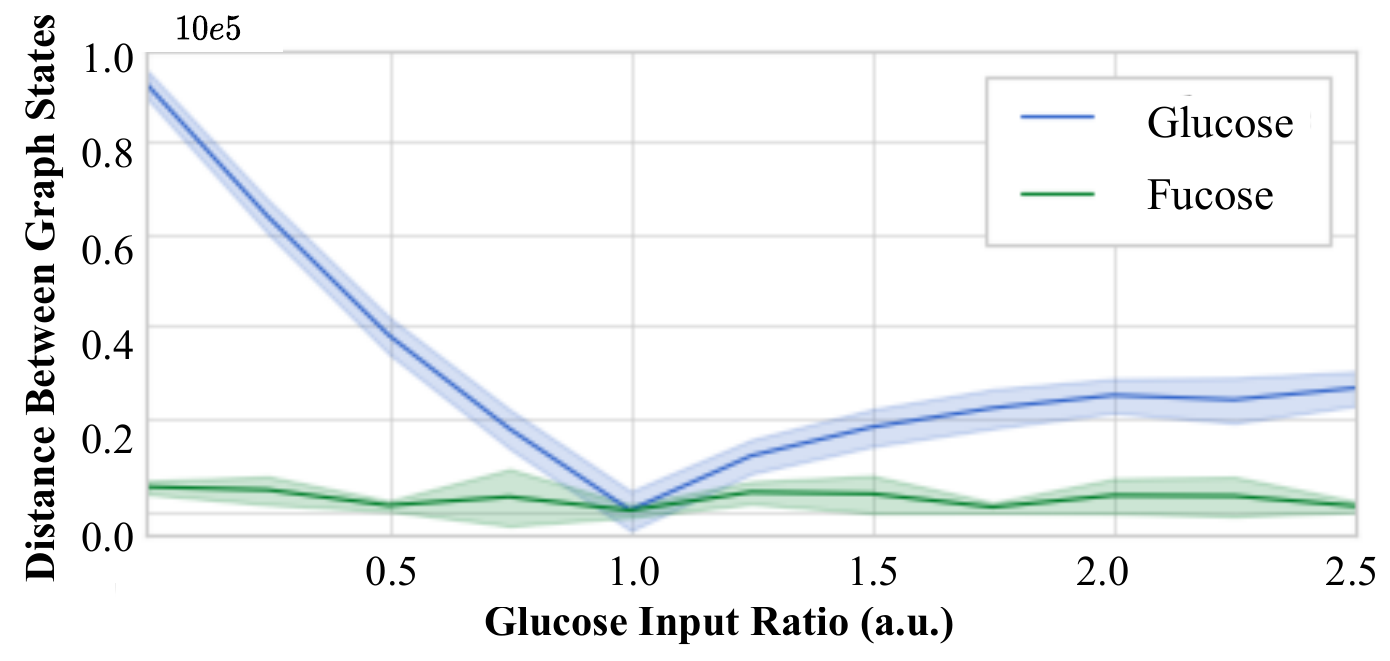}
    \caption{Behaviors of overall graph weights against the changes in inputs and their concentrations.}
    \label{fig:InputVsGraphDistance}
\end{figure}
\sm{Here}, we present the results for the analyses mentioned in Section \ref{sec: InputVsStructure}. The analyses are conducted by regulating the inputs glucose rate $C^r_{(host, Mem)}(M_{glu})$ and fucose rate $C^r_{(host, Mem)}(M_{fse})$ \sm{from the host cells} to the system that contains memory of existing metabolites and evaluating the  \sm{human GB} compositional changes. The simulation for these experiments only contains growth dynamics of \textit{Faecalibacterium}, \textit{Eubacterium} and \textit{Escherichia} bacteria. 

Figure \ref{fig:A1-GlucoseVsGrowth} illustrates the impact of glucose on the three bacterial populations. Figure \ref{fig:A1-GlucoseSubGraph} shows the glucose consumption graph $\Gamma_{glu}$, which is a subgraph of SCFA production graph, $\Gamma_{glu} \subseteq \Gamma_{SCFA}$. The colors used in Figures \ref{fig:A1-GlucoseMiddleLayer.pdf} and \ref{fig:A1-GlucoseMCLayer.pdf} follow the same color scheme as in Figure \ref{fig:A1-GlucoseSubGraph}. Figure \ref{fig:A1-GlucoseMiddleLayer.pdf} shows the behaviors of edge weight deviations of the bacteria in the host due to the changes in the glucose input rates $C^r_{(host, Mem)}(M_{glu})$. Meanwhile, Figure \ref{fig:A1-GlucoseMCLayer.pdf} depicts the variation of population sizes due to the changes in $C^r_{(host, Mem)}(M_{glu})$ as a fraction of the value when the input is at the standard level. The variations of the input rate $C^r_{(host, Mem)}(M_{glu})$ alters the intermediate interaction from any bacterial population $B_k$ to other species $B_{k'}$ through acetate, $C_{(B_k, B_{k'})}(M_{ace})$ and lactate $C_{(B_k, B_{k'})}(M_{lact})$, which are required for the growth of \textit{Faecalibacterium} and \textit{Eubacterium}, respectively. Figure \ref{fig:A1-GlucoseMiddleLayer.pdf} explains the graph theoretical behavior of indirect influence on the growth dynamics of the respective bacterial populations. The growth of \textit{Eubacterium} keeps increasing steadily until the $C^r_{(host, Mem)}(M_{glu})$ is twice of the standard level, while the growths of the other two bacterial populations converges to the static standard level. This phenomenon is manifested by the stoichiometry of the metabolite conversion, where an acetate molecule is produced by one glucose molecule while in the case of the lactate molecule it requires two glucose molecules. The growth of \textit{Escherichia} is directly altered by the variations of glucose inputs and the behaviors is similar to that of the \textit{Faecalibacterium} population. We calculated the Mean Standard Error (MSE) for the experiment by iterating the experiment 20 times and the maximum MSEs recorded for any metabolite is 0.03374.

\begin{figure*}[t!]
\centering
\subfloat[\label{fig:A2-Bacteroides-SNR}]{
\includegraphics[trim=0 0 0 0,clip,width=0.33\textwidth]{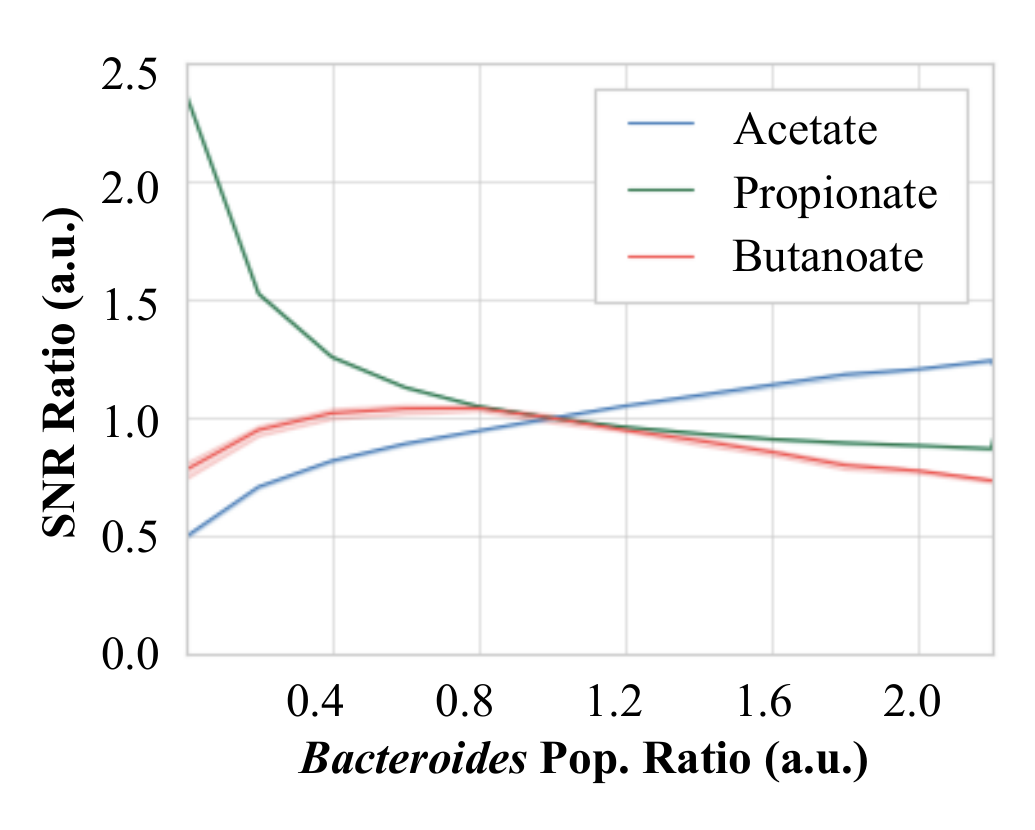}}
\subfloat[\label{fig:A2-Faecalibacterium-SNR}]{
\includegraphics[width=0.33\textwidth]{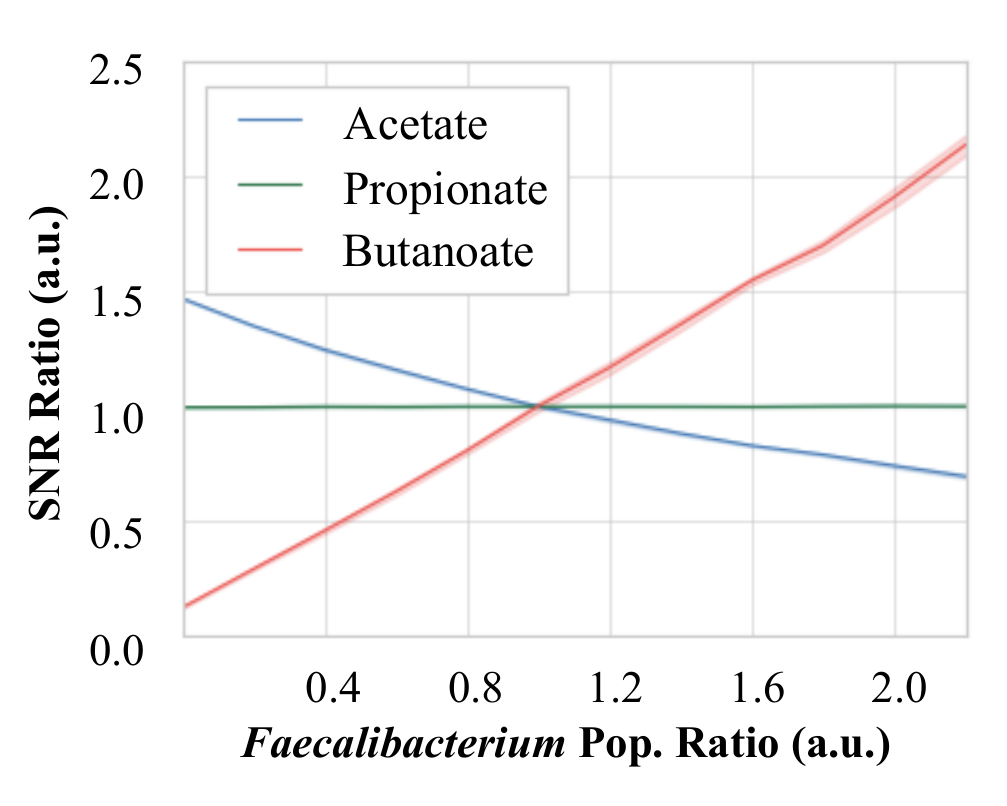}}
\subfloat[\label{fig:A2-Ruminococcus-SNR}]{
\includegraphics[width=0.33\textwidth]{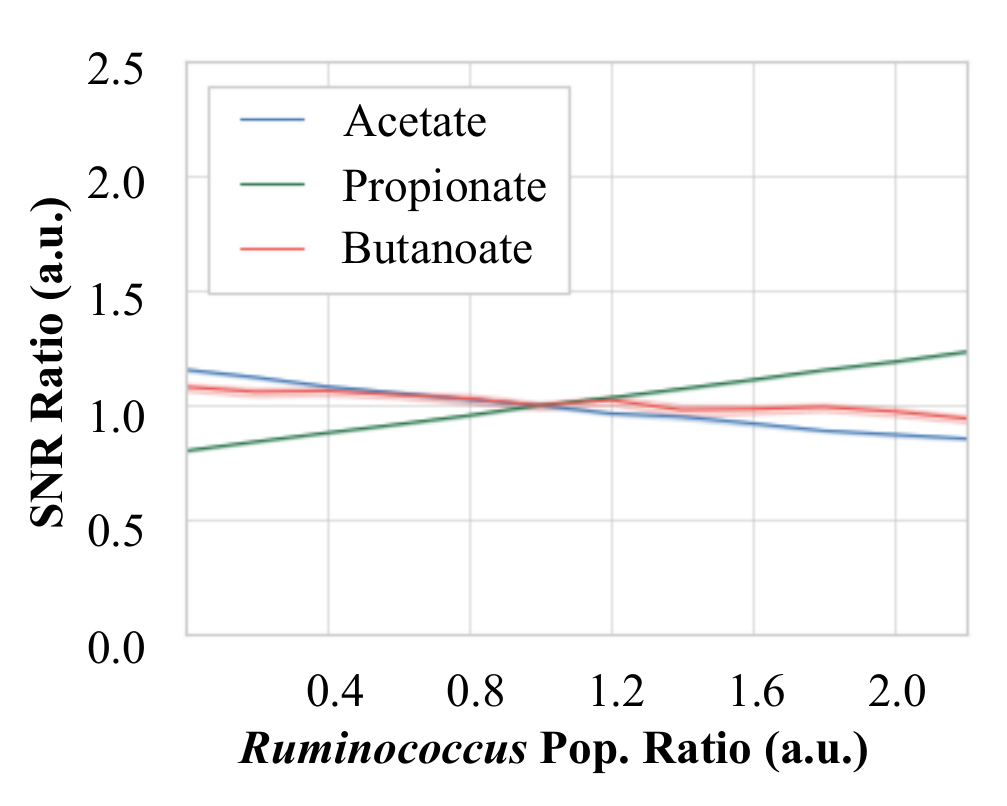}}

\caption{Simulation results for SNR of three output signal with the changes in population sizes of \textit{Bacteroides}, Feacalibacterium and \textit{Ruminococcus}. The populations sizes are changed as a ratio of the standard level: (a) SNR results for \textit{Bacteroides} population, (b) SNR results for Feacalibacterium population, and (c) SNR results for \textit{Ruminococcus} population.}
\label{fig:A2-BacteriaVsSNR}
\end{figure*}

\begin{figure*}[t!]
    \centering
    \includegraphics[trim=0 0 80 0,clip,width=0.9\textwidth]{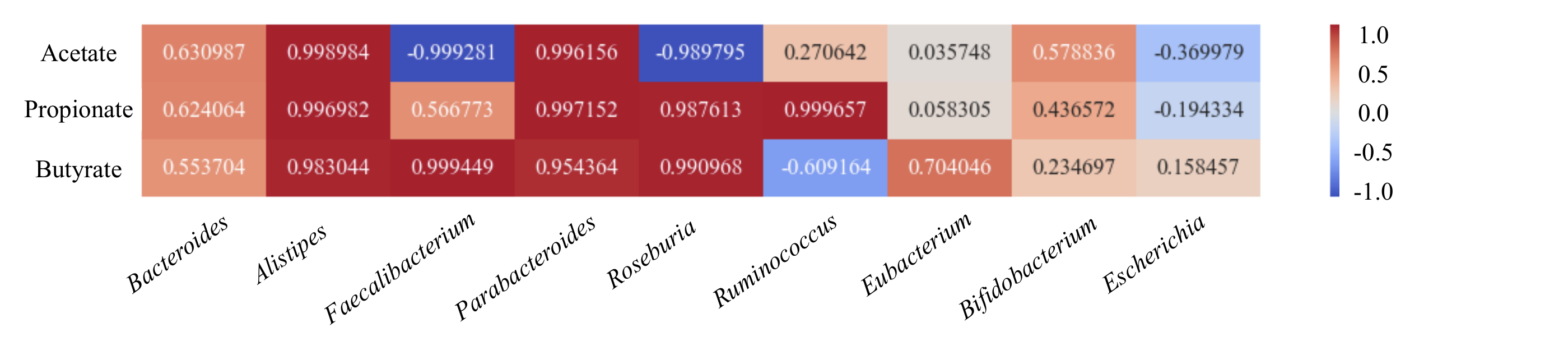}
    \caption{Pearson correlation heat map of the impact on the three output signals by nine bacterial populations.}
    \label{fig:An2-Correlation}
\end{figure*}

Deviations of a bacterial population concentration refer to deviations in node weights according to the (\ref{eq:NodeSPP}) and (\ref{eq:node_weight}). Figure \ref{fig:Weights} represents the node weight deviation compared to standard graph state $S_{0}$ due to the variability in inputs. This analysis reveals the impact of different input conditions on the molecular signal performance $SPP$ of bacterial populations.

While Figure \ref{fig:Weights} explains the node weight variations, Figure \ref{fig:InputVsGraphDistance} focuses on the overall interaction weight behaviors compared to $S_{0}$. This graph provides an insight into how the structure is being modified by the input variability. When the $C^r_{(host, Mem)}(M_{glu})$ is low compared to the standard level, the graph deviates significantly from the standard level and when the $C^r_{(host, Mem)}(M_{glu})$ exceeds the standard level, the graph starts to deviate again from the standard structure, but with a lesser magnitude compared to a weaker signal (the standard level is 1.0). This reveals that the human GB is more sensitive to low glucose  concentrations. The experiment is repeated for the fucose input rates $C^r_{(host, Mem)}(M_{fse})$ as well, but the impact is minimal compared to $C^r_{(host, Mem)}(M_{glu})$.

\subsection{Analysis 2: Human GB structure effect on the graph outputs}
\label{sec:StructureVSOutput}
In this section, we analyze the direct and indirect impacts of the human GB compositional changes on the network behaviors. The analyses are conducted by altering the bacterial population sizes manually on the virtual GB and extracting the metabolite production data with respect to each alteration. The resulting behaviors of the MC layer are explained using the graph analyses. Although we conduct similar experiments for all the nine populations, we only show results on \textit{Bacteroides}, \textit{Faecalibacterium} and \textit{Ruminococcus} populations as they provide a better understanding of the metabolite production dynamics of the human GB. Figures \ref{fig:A2-BacteroideVsSCFA}, \ref{fig:A2-FaecaliVsSCFA} and \ref{fig:A2-RuminococcusVSSCFA} illustrate the influence of \textit{Bacteroides}, \textit{Faecalibacterium} and \textit{Roseburia} populations on the SCFA production. 

Figure \ref{fig:A2-BacteroideVsSCFA} shows the impact of \textit{Bacteroides} population size variation on the human GB SCFA production. In this experiment, we focus on the graph $\Gamma_{Bct}$ considering only the interactions that are related to the \textit{Bacteroides} population, as shown in Figure \ref{fig:A2-BacteroidesSubGraph}. $\Gamma_{Bct}$ is a subgraph of SCFA production graph, $\Gamma_{Bct} \subseteq \Gamma_{SCFA}$. The color scheme used in Figures \ref{fig:A2-BacteroidesMiddleLayer} and \ref{fig:A2-BacteroidesMCLayer} follow the same color scheme as in Figure \ref{fig:A2-BacteroidesSubGraph}. The metabolite inputs to the graph and the population sizes are maintained fixed at the standard level except for the \textit{Bacteroides} population size. We modify the population size of \textit{Bacteroides} ($|B_{Bct}|$) from zero cells to 2.2 times the standard population size. Figure \ref{fig:A2-BacteroidesMiddleLayer} explains the behaviors of the intermediate links from \textit{Bacteroides} to \textit{Faecalibacterium} node through acetate $C_{(Bact, Fae)}(M_{ace})$, \textit{Bacteroides} to \textit{Eubacterium} populations through lactate $C_{(Bact, Eub)}(M_{lact})$ and \textit{Bacteroides} to \textit{Roseburia} populations through acetate $C_{(Bact, Ros)}(M_{ace})$ while Figure \ref{fig:A2-BacteroidesMCLayer} shows SCFA production behaviors in the MC layer due to changes in the \textit{} population size.
From Figure \ref{fig:A2-BacteroidesMCLayer}, it is evident that all the SCFAs have strong positive relationships with the population size of \textit{Bacteroides}. Acetate and propionate are direct productions of \textit{Bacteroides} cells. As a result of that, acetate and propionate outputs show steady trends against the increment of \textit{Bacteroides} population sizes. Moreover, the edge weight variations shown in Figure \ref{fig:A2-BacteroidesMiddleLayer} justify the butyrate signal behavior in the MC layer shown in Figure \ref{fig:A2-BacteroidesMCLayer}. To be more precise, the butyrate output curve starts to become flatter when the \textit{Bacteroides} population size $|B_{Bct}|$ is greater than 0.8 times the standard value.  The graph theoretical quantification of links also shows the same trend in Figure \ref{fig:A2-BacteroidesMiddleLayer}, emphasizing that the graph theoretical measures can be used to explain the metabolite production behaviors.

In the same way, Figure \ref{fig:A2-FaecaliVsSCFA} illustrates the results for a similar experiment on \textit{Faecalibacterium} population. Figures \ref{fig:A2-FaecalibacteriumSubGraph}, \ref{fig:A2-FaecalibacteriumMiddleLaye} and \ref{fig:A2-FaecalibacteriumMCLayer} represent the subgraph $\Gamma_{Fae}$ ($\Gamma_{Fae} \subseteq \Gamma_{SCFA}$), edge  and the MC layer behaviors, respectively. Similarly to the previous analysis, we modify the population size of \textit{Faecalibacterium} $|B_{Fae}|$ ranging from zero cells to 2.2 times the standard population size. As the \textit{Faecalibacterium} cells consume acetate and produce butyrate, the rate of acetate consumption from the environment, $C^r_{(Mem, Fae)}(M_{ace})$ increases when the $|B_{Fae}|$ is increased. Hence, the weight of interaction between environment and \textit{Faecalibacterium} population, $C^w_{(Mem, Fae)}(M_{ace})$ increases which can be observed in Figure \ref{fig:A2-FaecalibacteriumMiddleLaye} and the resulting reduction in acetate output is visible in Figure \ref{fig:A2-FaecalibacteriumMCLayer}. Moreover, since \textit{Faecalibacterium} population is one of the key butyrate producers, there is a clear positive relationship evident between $|B_{Fae}|$ and butyrate. Due to the smaller population size of \textit{Roseburia} population, the influence on the metabolite production is relatively low, which can be observed from Figure \ref{fig:A2-FaecalibacteriumMiddleLaye}. For all the graphs the maximum MSEs are calculated below 0.03087.

Figure \ref{fig:A2-RuminococcusVSSCFA} elaborates the analytical results for a similar analysis on \textit{Ruminococcus} population size variation as we conduct on \textit{Bacteroides} and \textit{Faecalibacterium} populations. First, we extract a graph considering the metabolite consumption and production of \textit{Ruminococcus} cells $\Gamma_{Rcc}$ which is a subgraph of the SCFA production graph, $\Gamma_{Rcc} \subseteq \Gamma_{SCFA}$. Figures \ref{fig:A2-RuminococcusSubGraph}, \ref{fig:A2-RuminococcusMiddleLayer} and \ref{fig:A2-RuminococcusMCLayer} represent the subgraph $\Gamma_{Rcc}$ edge behaviors and the metabolite outputs, respectively. Similar to previous analysis, we modify the population size of \textit{Ruminococcus} $|B_{Rcc}|$ ranging from 0 - 2.2 times the standard population size. In this analysis, we considered fucose as the input signal to the \textit{Ruminococcus} population, $C_{(host,Rum)}(M_{fse})$ due to the metabolic switching from converting glucose into acetate to converting fucose into propionate of \textit{Ruminococcus} cells in the presence of fucose. It is clear in Figure \ref{fig:A2-RuminococcusMiddleLayer} the edge weight $C^w_{(host,Rum)}(M_{fse})$ is increased in parallel to the $|B_{Rcc}|$. Due to this behavior, the propionate production is increased, which can be observed in Figure \ref{fig:A2-RuminococcusMCLayer}. Again, these results confirm that the graph theoretical analysis reflects the behaviors of metabolite production in the MC Layer.

The MC layer results presented for the three analyses on \textit{Bacteroides}, \textit{Faecalibacterium} and \textit{Ruminococcus} populations (Figures \ref{fig:A2-BacteroidesMCLayer}, \ref{fig:A2-FaecalibacteriumMCLayer} and \ref{fig:A2-RuminococcusMCLayer}) are then interpreted in terms of SNR in Figure \ref{fig:A2-BacteriaVsSNR}. In the plots of this figure, SNR values are shown as ratios of the SNR value at the standard state of human GB and the bacterial population sizes are increased similar to the previous analyses. Here, we show the three SNRs of acetate, propionate and butyrate of three bacterial populations: \textit{Bacteroides}, \textit{Faecalibacterium} and \textit{Ruminococcus}. Figure \ref{fig:A2-Bacteroides-SNR} shows the SNR behaviors of the three SCFAs against the $|B_{Bct}|$. It is clearly evident that the acetate production is higher compared to the other two SCFAs when the $|B_{Bct}|$ is increased. This means, when the composition of human GB is changed as the $|B_{Bct}|$ increases, the output of the GB also loses the balance and tends to produce more acetate compared to the other two SCFAs. On the contrary, propionate production rate reduces when the $|B_{Bct}|$ increases. When the population size of \textit{Bacteroides} $|B_{Bct}|$ is smaller than the standard level, the system tends to produce molecular signal with higher deviated ratios, but when $|B_{Bct}|$ is greater than the standard level, the deviation is relatively low.
Figure \ref{fig:A2-Faecalibacterium-SNR} shows the SNR behaviors of the three SCFAs against the $|B_{Fae}|$. Since \textit{Faecalibacterium} is the main butyrate producer of this network, the butyrate SNR increases with the $|B_{Fae}|$ increment. Hence, compositional imbalance related to \textit{Faecalibacterium} causes a significant imbalance in output molecular signal ratios. Furthermore, due to the acetate consumption of \textit{Faecalibacterium}, the acetate signal becomes weaker, resulting in the acetate SNR deviating from the standard level. Moreover, Figure \ref{fig:A2-Ruminococcus-SNR} elaborates the SNR behavior  due to variations in the \textit{Ruminococcus} population size, $|B_{Rcc}|$. Since, the population size of \textit{Ruminococcus} is small, the impact of it on the output signal ratio is relatively low compared to the other populations. The SNR for propionate is clearly increased with the \textit{Ruminococcus} population size increases as \textit{Ruminococcus} is a propionate producer.

The heat map shown in Figure \ref{fig:An2-Correlation} explains the correlation between each bacterial population and the SCFA abundance in the gut environment. Although the \textit{Bacteroides} is the biggest producer of all the SCFAs, it has a week correlation with SCFAs compared to other producers such as \textit{Alistipes} and \textit{Parabacteroides}. This reveals that the reduction of glucose consumption by \textit{Bacteroides} increases the other bacterial population resulting in boosted SCFA production. Note that, even the SCFA production of the other bacterial population is boosted in the absence of \textit{Bacteroides}, the overall production is lower. Since the \textit{Faecalibacterium} and \textit{Roseburia} consume acetate, the heat map shows a strong negative correlation with acetate. Interestingly, this heat map indicates the metabolic switching for \textit{Escherichia} which is switching from a SCFA producer to consumer of high acetate concentrations. This is same for the \textit{Ruminococcus} when the fucose concentration is not sufficient for the increased population, it switches from fucose consumption to glucose consumption reducing the intermediate metabolite production which causes a reduction in butyrate production. 

\section{Conclusion}\label{sec:conclusions}
The gut bacteriome has been largely investigated due to its importance to the human health. We contribute to this research topic by introducing a two-layer GB interaction model to investigate the impacts of bacterial population compositional changes on the overall structure of the human GB utilizing data collected from MicrobiomeDB and NJS16 databases. Our proposed human GB interaction model combines a bacterial population graph layer, which models the structure typically found in the human GB (i.e. bacterial populations genus and sizes), with a molecular communications layer, which models the exchange of metabolites by the bacterial populations in this structure. Supported by these models we also developed a virtual GB to simulate the metabolic interactions that typically occurs in the human GB. These simulations allowed us to study the impacts caused by the metabolite exchanges on the human GB structure (i.e. nodes weight and hamming distance). Through our analyses we found that the molecular input availability affect differently the bacterial populations in the human GB by modifying the nodes and edges weights of our GB interaction model. Our results also show that modifications in the human GB structure, in specific changing the sizes of \textit{Bacteroides}, \textit{Faecalibacterium} and \textit{Ruminococcus} populations can lead to improvement/reduction on the production of SCFA, which may result in metabolic diseases in humans. Based on our results we also infer that there is an intrinsic
relationship between the investigated bacterial populations sizes, the increase/decrease of specific metabolites (acetate, butyrate and propionate) and the overall balance of the human GB. These results can support the development of novel strategies to treat unbalanced human GB and can provide insights on the role of other metabolites and molecules on the maintenance of a healthy gut bacteriome.

\appendix

\begin{figure*}
\centering
\begin{subfigure}[b]{\linewidth}
\includegraphics[width=0.96\linewidth]{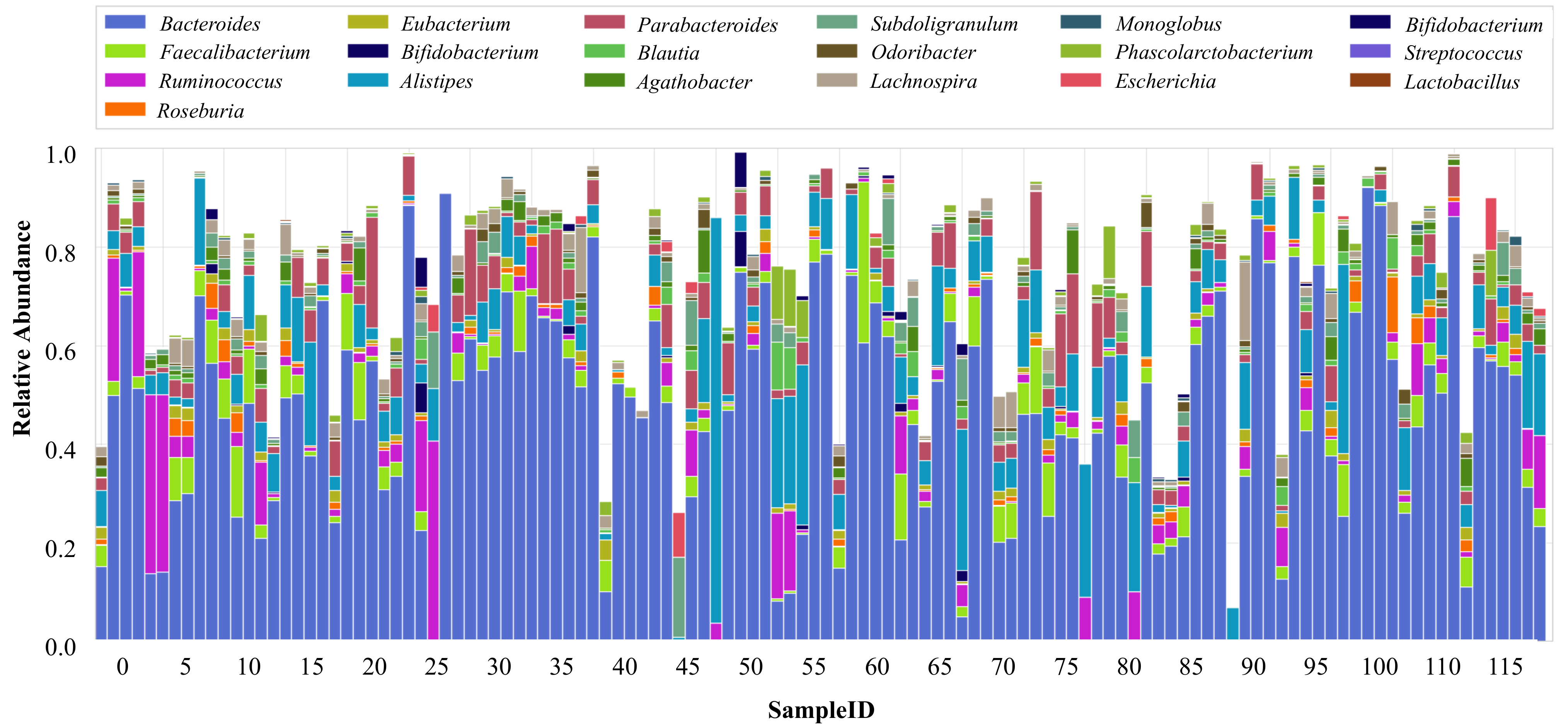}
\caption{Human GB data.}
\end{subfigure}

\begin{subfigure}[b]{\linewidth}
\includegraphics[width=0.96\linewidth]{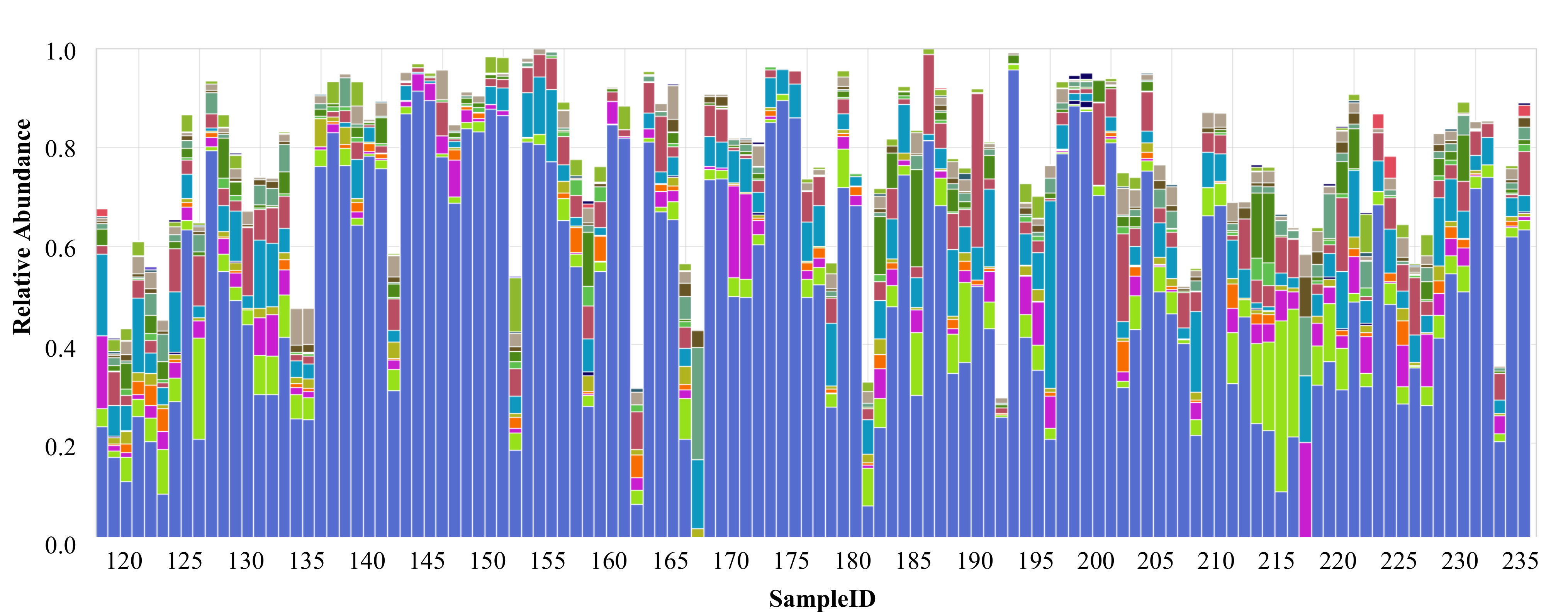}
\caption{Human GB data (cont.).}
\end{subfigure}
\begin{subfigure}[b]{\linewidth}
\includegraphics[width=0.96\linewidth]{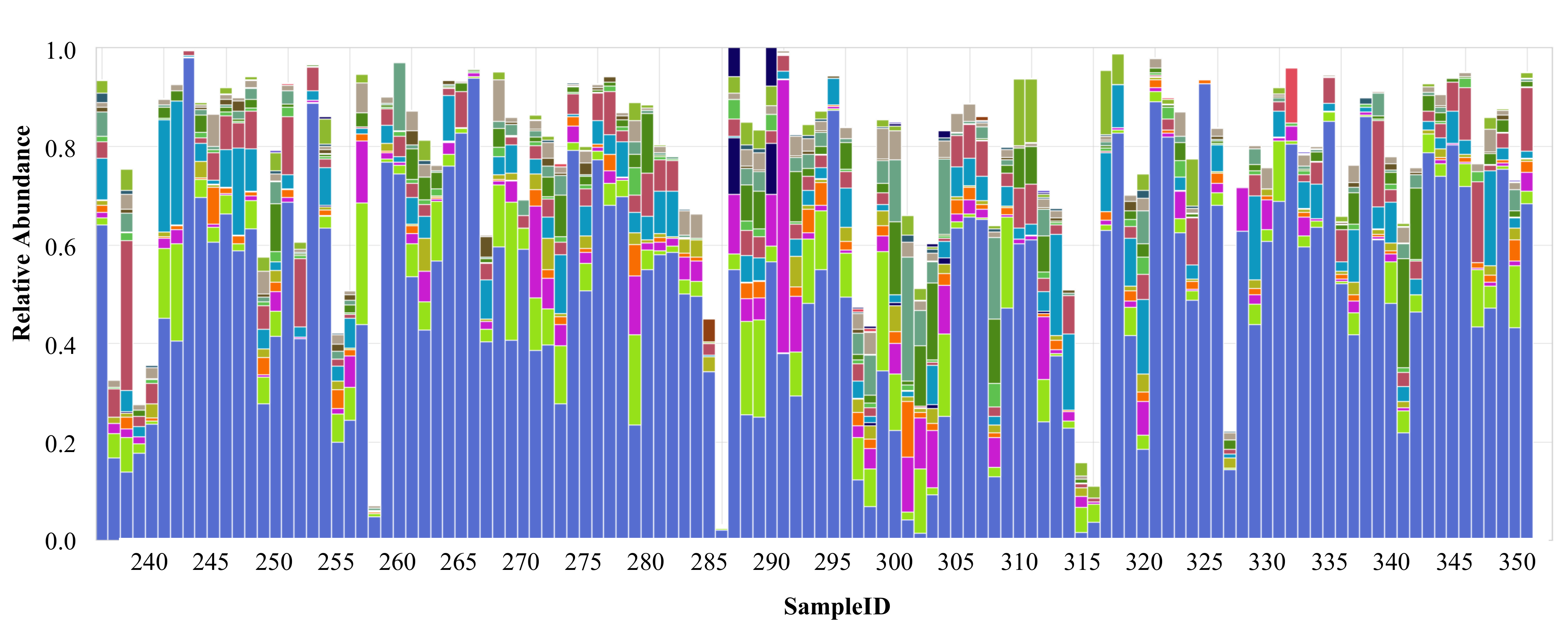}
\caption{Human GB data (end).}
\end{subfigure}

\caption{Relative abundances of 352 gut bacteriome samples.}
\label{fig:Composition}
\end{figure*}

Here, we illustrate the RA data used in the case study for SCFA production in the human GB. This data was collected from the MicrobiomeDB, and it is presented in Figure \ref{fig:Composition}. Please note that these figures shows the collective species RA for that particular genus. 

\section*{Acknowledgement}
This publication has emanated from research conducted with the financial support of Science Foundation Ireland (SFI) and the Department of Agriculture, Food and Marine on behalf of the Government of Ireland under Grant Number [16/RC/3835].

\ifCLASSOPTIONcaptionsoff
  \newpage
\fi

\bibliographystyle{ieeetr}
\bibliography{sample}

\end{document}